\documentclass[%
preprint,
 amsmath,amssymb,
 aps,
]{revtex4-2}
\usepackage{graphicx}
\usepackage{epstopdf}
\usepackage{color}
\usepackage{afterpage}
\usepackage{dcolumn}
\usepackage{bm}
\newcommand{\solidcirc}{---$\circ$---}

\newcommand{\solidtri}{---$\vartriangle$---}
\newcommand{\solidtridown}{---$\triangledown$---}

\newcommand{\dashed}{\hbox{{--}\,{--}\,{--}\,{--}}}
\newcommand{\chndot}{---\,$\cdot$\,---}

\newcommand{\mylab}[3]{\raisebox{#2}[0mm][0mm]{%
		\makebox[0mm][l]{\hspace*{#1}\textbf{#3}}}}
\def\bra{\langle}
\def\ket{\rangle}

\def\beq{\begin{equation}}
\def\eeq{\end{equation}}

\def\rev#1{{\color{black}#1}}
\def\notyet#1{{\color{black}#1}}

\begin{document}

\preprint{APS/123-QED}

\title{Outer scaling of self-similar adverse-pressure-gradient \\ turbulent boundary layers}

\author{Atsushi~Sekimoto}
\email{asekimoto@cheng.es.osaka-u.ac.jp}
\affiliation{Department of Materials Engineering Science, Graduate School of Engineering Science, Osaka University, Toyonakaa, Osaka 560-8531, Japan}

\author{Vassili~Kitsios}%
\affiliation{%
CSIRO Oceans and Atmosphere, Castray Esplanade, Battery Point, TAS 7004, Australia
}%
\altaffiliation[Also at ]{Laboratory for Turbulence Research in Aerospace and Combustion, Department of Mechanical and Aerospace Engineering, Monash University, Clayton Campus, Melbourne, VIC 3800, Australia}
%

\author{Callum Atkinson}
\author{Julio Soria}
\affiliation{
Laboratory for Turbulence Research in Aerospace and Combustion, Department of Mechanical and Aerospace Engineering, Monash University, Clayton Campus, Melbourne, VIC 3800, Australia
}%

\date{\today}

\begin{abstract}
The prediction of turbulent boundary layer (TBL) flow over a convex surface as in aircraft wings or gas turbine blades is a challenging problem. Finding a universal scaling law of turbulence statistics of TBLs over a wide range of adverse pressure gradients (APG) remains unresolved.
  Here, we introduce characteristic length and velocity scales for APG-TBLs and nondimensionalise the turbulence statistics of the recent canonical self-similar APG-TBLs by Kitsios {\it et al.} ({\it J. Fluid Mech.}, vol.829, 2018, pp. 392--419).
  The characteristic length scale, which is termed the `shear thickness', $\delta^\ast$, is defined as the location which corresponds to the end of an actively sheared region 
  in a turbulent shear flow, where the nondimensional shear rate normalised by the kinetic energy and the dissipation rate is approximately constant.  
 Next, we show an universal scaling using a mixed velocity, termed the `friction-pressure velocity', $u^\ast$, which is based on total shear stress. 
 It is revealed that the velocity fluctuations and the Reynolds stresses in TBLs over a wide range of APGs agree well with those in TBLs with  zero-pressure-gradient (ZPG). 
The present scaling is used to scale the kinetic energy balance in TBLs, and compare them to other shear flows. Furthermore, a scaling for small-scale properties, i.e. vorticities, using $\delta^\ast$ and $u^\ast$ is also obtained assuming the local equilibrium in the inertial range. 
  The present scaling for wall-bounded shear flows, including TBLs over a wide range of pressure gradients, implies that the underlying instantaneous turbulence structures have common features under a proper scaling and is key to the development and application of turbulent models.
\end{abstract}

\maketitle
%
%

\section{Introduction}\label{sec:Intro}
\notyet{
One of the most accepted universal scalings is the logarithmic law for the mean velocity in the overlap region.
The applicability of the log-law is, however, limited to simple flows, and a more universal scaling law for the statistics of wall-bounded turbulent shear flows has not yet been developed. 
Here we aim to extend such scaling laws for flows applicable to 
flows  such as for example, the turbulent wall-bounded flows over 
wings of aircraft, diffusers, and blades of gas turbines.
}
The prediction of flow separation of adverse-pressure-gradient turbulent boundary layer (APG-TBL) is a challenging problem and many aspects of turbulent structures and the scaling of turbulence statistics in APG-TBL remain unresolved.  
The canonical APG-TBL at the verge of separation is self-similar, in the sense that each of the terms in the governing equations have the same proportionality with the streamwise position~\cite{Townsend1976book,Townsend1960,MellorGibson1966}. 
Self-similar APG-TBLs have been approximately achieved in the recent direct numerical simulations (DNS) \cite{Skote1998, KitsiosAtkinsonSilleroBorrellGungorJimenezSoria2016,KitsiosSekimotoAtkinsonSilleroBorrellGungorJimenezSoria2017} up to $\beta \approx 39$, where $\beta = (\delta_1/\tau_\mathrm{w}) P_e^\prime$ is the Clauser's pressure gradient factor, $\delta_1$ is the displacement thickness, $\tau_\mathrm{w}$ is the mean wall shear stress, $P_e^\prime$ is the pressure gradient, and $P_e$ is the far-field streamwise dependent pressure per an unit density. 
They confirmed that the outer peaks of the velocity fluctuations show self-similar collapse when using the characteristic length and velocity scale of the displacement (or momentum) thickness and the external velocity, respectively, since the upstream history effect is minimised and the condition of self-similarity are satisfied within the domain of interest.

The lack of an universal definition for the characteristic thickness of APG-TBLs complicates the comparison of different studies.
\citet{Lighthill1963book} and \citet{SpalartWatmuff1993} used the displacement thickness using the vorticity form in order to take into consideration the streamwise gradient of the mean vertical velocity, which is not negligible under a strong pressure gradient.
However, the displacement thickness, $\delta_1$, does not work as a common characteristic outer length scale of APG-TBLs over a wide range of $\beta$, since the ratio between the TBL thickness, for example $\delta_{99}$ (the distance from the wall at which the mean velocity recovers 99\% of the free-stream), and $\delta_1$ depends highly on $\beta$.
\citet{VinuesaBobkeOrluSchlatter2016} mensioned that $\delta_{99}$ does not represent the boundary layer thickness for $\beta > 7$.
\citet{AlfredssonOrlu2010} defined the edge of TBLs using the diagnostic plot concept, 
where 2\% of the mean streamwise velocity scaled by the shape factor is used.
 \citet{GungorMacielSimensSoria2016} have tested a mixing-layer-like scaling for a large-defect APG-TBL. \citet{KitsiosSekimotoAtkinsonSilleroBorrellGungorJimenezSoria2017} define the edge of the boundary layer  as the location where the mean vorticity is 0.2\% of the mean vorticity at the wall.
None of the definitions of the boundary layer thickness, $\delta$, are free from an arbitrary choice of a threshold.
It is also worth noting that a constant free-stream velocity is not achieved in any TBL, particularly in the case of an APG-TBL, and 
the concept that a constant free-stream velocity exists is not appropriate. The measurement point of the reference free-stream velocity somewhat affects the definition of $\delta_{99}$. In this sense, the thicknesses of all TBLs are not consistent.
%

With respect to the scaling for velocities, the friction velocity, $u_\tau$, and length, $\nu/u_\tau$, where $\nu$ is kinematic viscosity, are 
used as `wall-unit' scaling for wall-bounded turbulence, and an universal logarithmic
(or exponential) profile of the mean streamwise velocity is observed. 
Such an universal wall function in APG-TBLs is not fully understood, and a customised wall function for each $\beta$ is used.  
\citet{ZagarolaSmitsFEDSM1998} have shown a velocity scale for the mean velocity deficit. However, it is not free from the definition of $\delta$. The effect of the pressure gradient, $P_e^\prime$, should be taken into account for the proper scaling of statistics of APG-TBLs over a wide range of $\beta$. 
One of the candidates for the characteristic velocity is 
the pressure velocity, $u_P = \sqrt{ \delta P_e^\prime}$, whose definition requires a characteristic thickness, $\delta$. 
A proper definition of $\delta$ paves a way for an universal scaling for the velocity fluctuations in APG-TBLs using the friction and pressure velocity.
\citet{SkoteHenningson2002} derived a mixed velocity scale, 
$ u^{\ast 2} \equiv u_\tau^2 + u_\nu^3 (y /\nu)$
and corresponding wall-normal coordinate $y^\ast=\nu/u^\ast$ with their definition based on the $u_\nu \equiv (\nu P_e^\prime)^{1/3}$ as in \citet{TennekesLumley1972book} and \citet{Townsend1976book}. 
The mean velocity in the viscous region was successfully scaled by the mixed velocity. 
However, how the velocity fluctuations scale using the mixed velocity over a wide range of APG-TBLs has not been shown, possibly because of the lack of universal definition for characteristic outer-length scale over a wide range of APG-TBLs.
\citet{SekimotoAtkinsonSoria2018} investigated the pressure gradient effects 
in a minimal-span Couette-Poiseuille system, where the large-scale length is limited by the spanwise box dimension, and this parallel flow can be considered as a model problem of the APG-TBL. 
It is revealed that the mixed-type velocity scale which is attributed to the local total shear stress scales the velocity fluctuations in the outer part of the parallel flow, implying that the mixed velocity is a good candidate to scale the turbulence statistics under the effect of adverse pressure gradient. 
In APG-TBLs, however, the outer-length scale should be rigorously determined in order to apply the mixed velocity scale.

In this study, we propose a new characteristic length scale of TBLs and show an universal scaling for the large- and small-scale properties, i.e. velocity fluctuations and enstrophy, in order to compare zero-pressure-gradient (ZPG-) and APG-TBLs consistently. 
\rev{For a canonical shear layer,} the Corrsin shear rate parameter~\citep{Corrsin1958} is given by 
\begin{equation}
  S_c \equiv \frac{\partial U(y) }{\partial y} \frac{q^2}{|\varepsilon|} \label{eq:S_c_intro}, 
\end{equation}
where $\partial U/\partial y$ is the mean velocity gradient, $q^2 = 2K$, $K$ is the mean kinetic energy and $\varepsilon$ is the pseudo-dissipation rate. 
$S_c$ is approximately constant in $y$ and has a similar value in the logarithmic layer of wall-bounded turbulence~\citep{Jimenez2013nearwall} and  the statistically-stationary homogeneous shear turbulence (SS-HST) \citep{SekimotoDongJimenez2016}.
Turbulent structures in such constant shear-rate regions of wall-bounded turbulent flows resemble unbounded homogeneous shear turbulence~\citep{DongLozanoSekimotoJimenez2017}.  
The new `shear thickness' presented in this study characterises 
the length scale of an actively sheared region. 

The small-scale properties (i.e. the vorticity magnitude with units of the inverse of time) can also be properly scaled by using the shear thickness and mixed velocity scale. 
Motivated by characterising turbulent/non-turbulent (TNT) interfaces, \citet{BorrelJimenez2016} proposed a self-similar scaling for the vorticity magnitude in ZPG-TBL, as $\omega^\prime \sim (\delta_{99}^+)^{-1/2} (u_\tau^2/\nu) = \sqrt{Re_\tau} (u_\tau/\delta_{99}) $, where $\delta_{99}^+ = Re_\tau$ is the boundary layer thickness scaled in wall units, equivalent to the friction Reynolds number, $Re_\tau$ based on $\delta_{99}$. 
In an APG-TBL at the verge of separation, however, such wall-unit scaling is undefined. 
A new scaling for vorticity in TBLs using the shear thickness and the friction and pressure velocities, therefore, needs to be developed in order to encompass all TBLs from $\beta=0$ to $\infty$. 

The paper is organised as follows. 
In the next section \S\ref{sec:nume}, the present database of a ZPG-, mild, strong APG-TBL~\citep{SilleroJimenezMoser2013,KitsiosAtkinsonSilleroBorrellGungorJimenezSoria2016,KitsiosSekimotoAtkinsonSilleroBorrellGungorJimenezSoria2017} 
as well as turbulent channel flow~\citep{LozanoJimenez2014pof} is summarised.  
The definition of the shear thickness is introduced in \S\ref{sec:shear-thickness}, and applied to the characterisation of the above database of wall-bounded turbulent shear flow. 
The velocity scales are introduced and tested in \S\ref{sec:scaling-vel}, and 
the proposed scaling is applied to the analysis of the mean velocity, Reynolds stresses, and the kinetic energy balance. In the end of \S\ref{sec:scaling-vel}, the self-similar scaling for vorticity magnitude is presented for APG-TBLs over a wide range of $\beta$, and the discussion with concluding remarks are presented in \S\ref{sec:discussion}.

\section{Database of wall-bounded shear flows}\label{sec:nume}
Table~\ref{tab:params} shows the details of the databases of wall-bounded turbulent flows used in this study. 
%
The axes in the streamwise, wall-normal and spanwise directions are $x$, $y$ and $z$. 
The corresponding velocity fluctuations with respect to the time-averaged mean ($U$, $V$, $W$) are ($u$, $v$, $w$), and the mean vorticity has only the spanwise component, $\Omega_z$, with $W=0$. Throughout the paper, $\bra \cdot \ket$ represents the average in time and space, i.e. the average in the homogeneous directions for channel flows, and the streamwise-dependent statistics of TBLs are averaged over the domain of interest where the flow has been shown approximately self-similar. 
The notation used for the derivative operator is $\partial_i$ where the subscript, $i=x$,~$y$,~$z$.
The direct numerical simulation (DNS) of TBLs is performed with a modified version of the code, {\tt OpenTBL}, developed by
\citet{SimensJimenezHoyasMizuno2009}, \citet{SilleroJimenezMoser2013} and \citet{BorrelSilleroJimenez2013}.
 A modified recycling method and far-field boundary conditions are used for the APG-TBLs~\citep{KitsiosAtkinsonSilleroBorrellGungorJimenezSoria2016,KitsiosSekimotoAtkinsonSilleroBorrellGungorJimenezSoria2017}.
Specifically, the mean pressure gradients for ZPG- and APG-TBLs are controlled by imposing an $x$-dependent wall-normal velocity under the irrotational boundary condition	 
at the boundary $y=L_y$, where $L_y$ is the height of the computational box.  
From the irrotational condition, the gradients $\partial_y U$ and $\partial_x V$ at $y=L_y$ are negative, yielding a turning point in the mean velocity profile for all TBLs, 
although the turning point is negligible for ZPG and becomes increasingly apparent as pressure gradient is increased.
%

It is appropriate to define the reference external velocity using the mean spanwise vorticity~\citep{Lighthill1963book,SpalartWatmuff1993},   
\begin{equation}\label{eq:defUe}
  U_e(x) \equiv U_\Omega(x, y_e), 
\end{equation}
where 
\begin{equation}\label{eq:defUomg}
  U_\Omega (x,y) \equiv - \int_0^{y} \Omega_z  (x, y^\prime) \mathrm{d} y^\prime,  
\end{equation}
and $y_e$ is defined as in \citet{KitsiosSekimotoAtkinsonSilleroBorrellGungorJimenezSoria2017} for their TBLs, 
and is replaced by the half channel width, $h$, for channel flows (CH42) of \citet{LozanoJimenez2014pof}, and by $\delta_{99}$ for the reference TBL (BL65) of \citet{SilleroJimenezMoser2013}.

The displacement and momentum thickness are defined, respectively, by 
\begin{eqnarray}
  \delta_1(x) &=& \frac{-1}{U_e} \int_0^{y_e} y \Omega_z(x,y) \mathrm{d} y, \\
  \delta_2(x) &=& \frac{-2}{U_e^2} \int_0^{y_e} y U_\Omega \Omega_z(x,y) \mathrm{d} y - \delta_1(x).
\end{eqnarray}
By using this velocity and outer scale, $U_e$ and $\delta_1$, respectively, the turbulence statistics of the mild and strong APG-TBL collapse within the domain of interest~\citep{KitsiosSekimotoAtkinsonSilleroBorrellGungorJimenezSoria2017}. 
Since $\beta$ drastically affects the mean velocity profile, $\delta_1$ cannot be used as a consistent characteristic length scale or thickness for all TBLs. Also, $\delta_{99}$ does not represent the boundary layer thickness for the strong APG case as also observed by \citet{VinuesaBobkeOrluSchlatter2016}.
Due to this, $\delta_{99}$ is not presented in Table~\ref{tab:params}.


\begin{table}
  \centering
  \begin{tabular}{ccccccc}
    \hline 
    Case & Lines \& symbols &  $y_e/\delta_1$  & $\delta^\ast/\delta_1$ &  $\beta^\ast$ & $Re_P^\ast$ & $Re_\tau^\ast$\\   
    \hline 
    strong APG  & \textcolor{red}{\solidtridown} & 4.8 & 1.6 &  59 & 4200 & 550  \\
    mild APG  & \textcolor{blue}{\solidtri} &  4.5 &  3.3  &  3.4 & 1100 & 610\\
    ZPG  & \textcolor{black}{\solidcirc} &  6.3 & 4.9 &  0.01  & 120 & 990 \\
    BL65 & \textcolor{magenta}{\chndot} &  6.1  & 5.1  &  0.005 & 120 & 1630 \\ 
    CH42 & \textcolor[rgb]{0,0.5,0}{\dashed} & 11.0  & 7.1 &  0 & 0 & 2700 \\
    \hline
  \end{tabular}
  \caption{The mean fraction of length scales for a ZPG-TBL ($\beta \approx 0$) and self-similar TBL with a mild ($\beta = 1$)~\citep{KitsiosAtkinsonSilleroBorrellGungorJimenezSoria2016} and strong APG ($\beta = 39$)~\citep{KitsiosSekimotoAtkinsonSilleroBorrellGungorJimenezSoria2017} in the self-similar domain of interest. 
    $\beta$ represents the pressure-gradient parameter~\citep{Clauser1954}. 
    $y_e$ is replaced by $h$ for channels at $Re_\tau = 4200$ (CH42) ~\citep{LozanoJimenez2014pof},
    and by $\delta_{99}$ for a ZPG-TBL at $Re_\theta=6500$ (BL65)~\citep{SilleroJimenezMoser2013}.
    $\beta^\ast$, $Re^\ast_P$ and $Re^\ast_\tau$ are the rescaled pressure parameter and the scaled pressure and friction Reynolds numbers, respectively. 
    }\label{tab:params}
\end{table}

\begin{figure}
	\centering
	\begin{minipage}{3ex}
		\rotatebox[origin=c]{90}{$(y/\delta_1) S_c$} 
	\end{minipage}
	\begin{minipage}{.43\linewidth}
		\includegraphics[width=1.0\linewidth,clip]
		{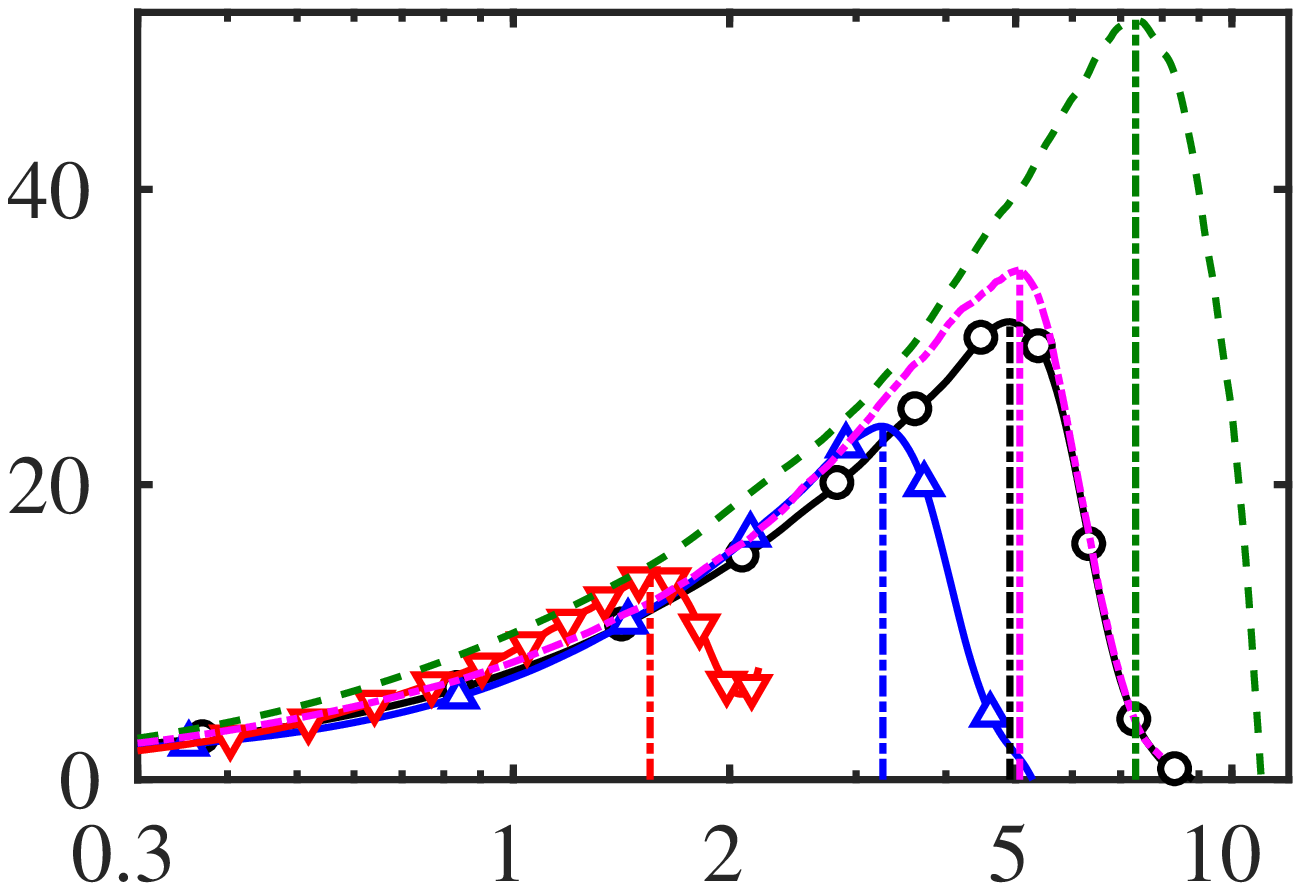}
		\centerline{$y/\delta_1$}
		\mylab{-0.3\linewidth}{0.7\linewidth}{(a)}
	\end{minipage}
	\hspace{3mm} 
	\begin{minipage}{3ex}
		\rotatebox[origin=c]{90}{$S_c$} 
	\end{minipage}
	\begin{minipage}{.44\linewidth}
		\includegraphics[width=1.0\linewidth,clip]
		{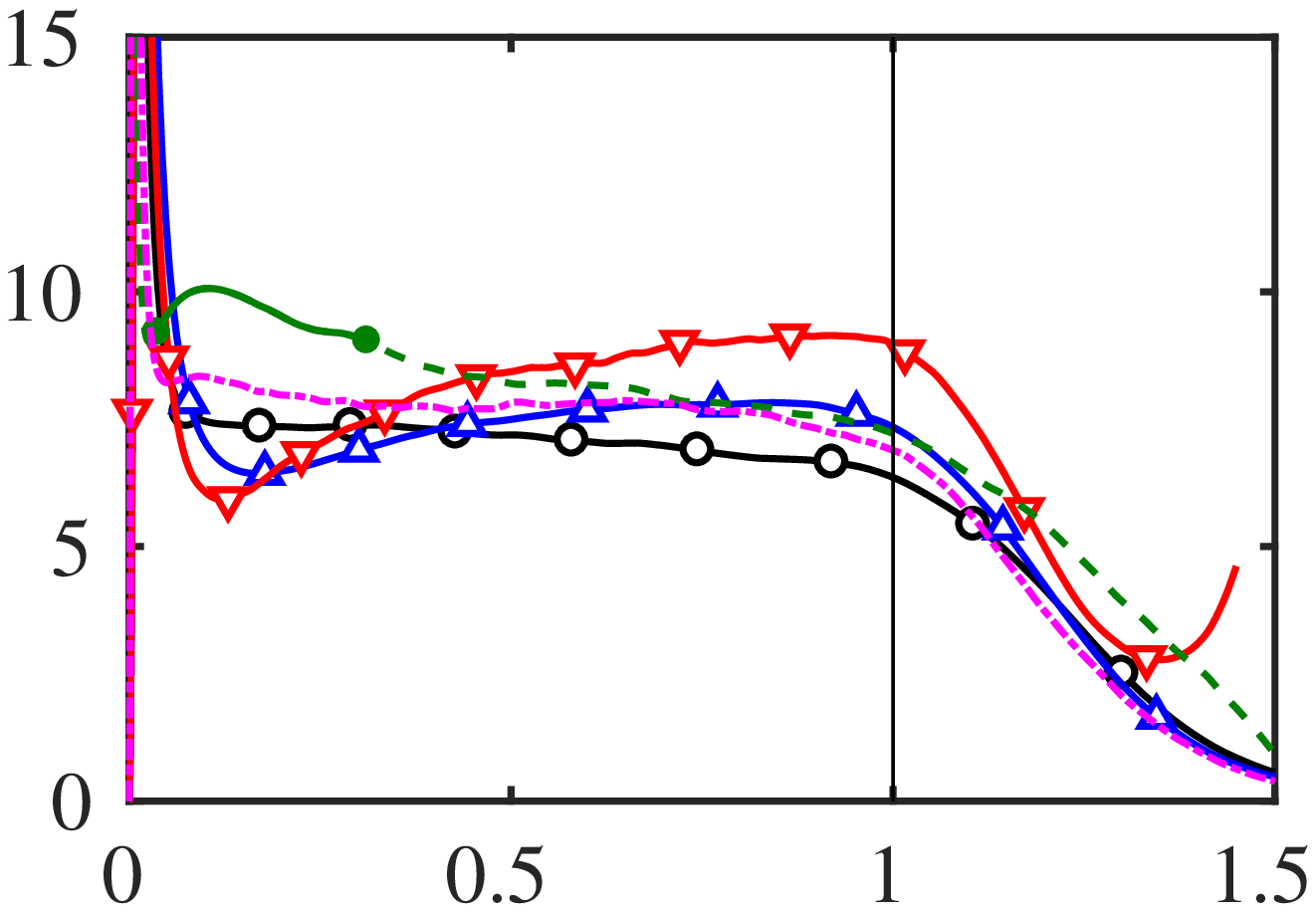}
		\centerline{$y/\delta^\ast$}
		\mylab{0.2\linewidth}{0.7\linewidth}{(b)}
	\end{minipage} 
	\\
	\begin{minipage}{3ex}
		\rotatebox[origin=c]{90}{$L_c/\delta^\ast$} 
	\end{minipage}
	\begin{minipage}{.44\linewidth}
		\includegraphics[width=1.0\linewidth,clip]
		{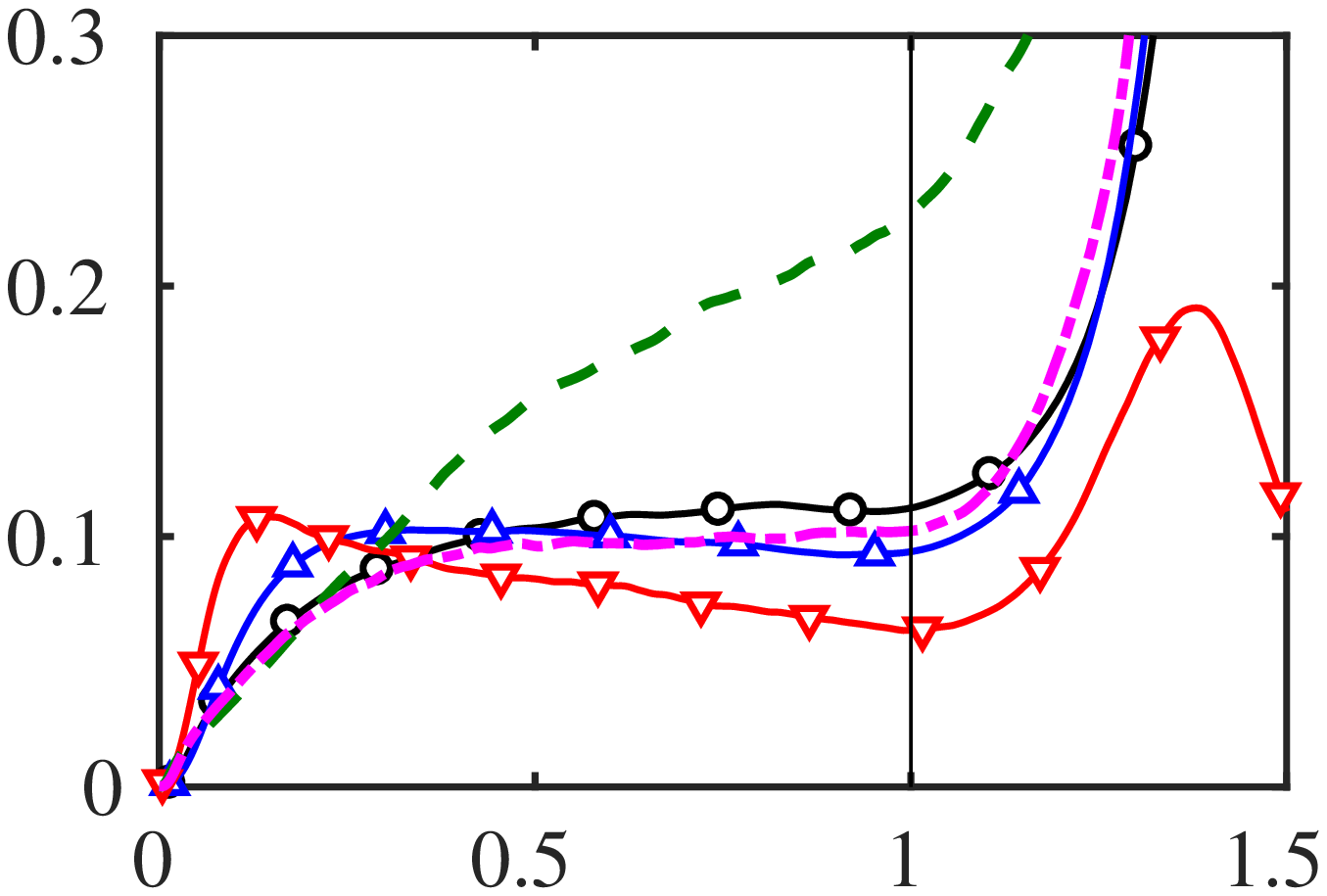}
		\centerline{$y/\delta^\ast$}
		\mylab{-0.3\linewidth}{0.7\linewidth}{(c)}
	\end{minipage}
	\hspace{3mm} 
	\begin{minipage}{3ex}
		\rotatebox[origin=c]{90}{$(L_c/\eta) Re_\lambda^{-3/2}$} 
	\end{minipage}
	\begin{minipage}{.4\linewidth}
		\includegraphics[width=1.0\linewidth,clip]
		{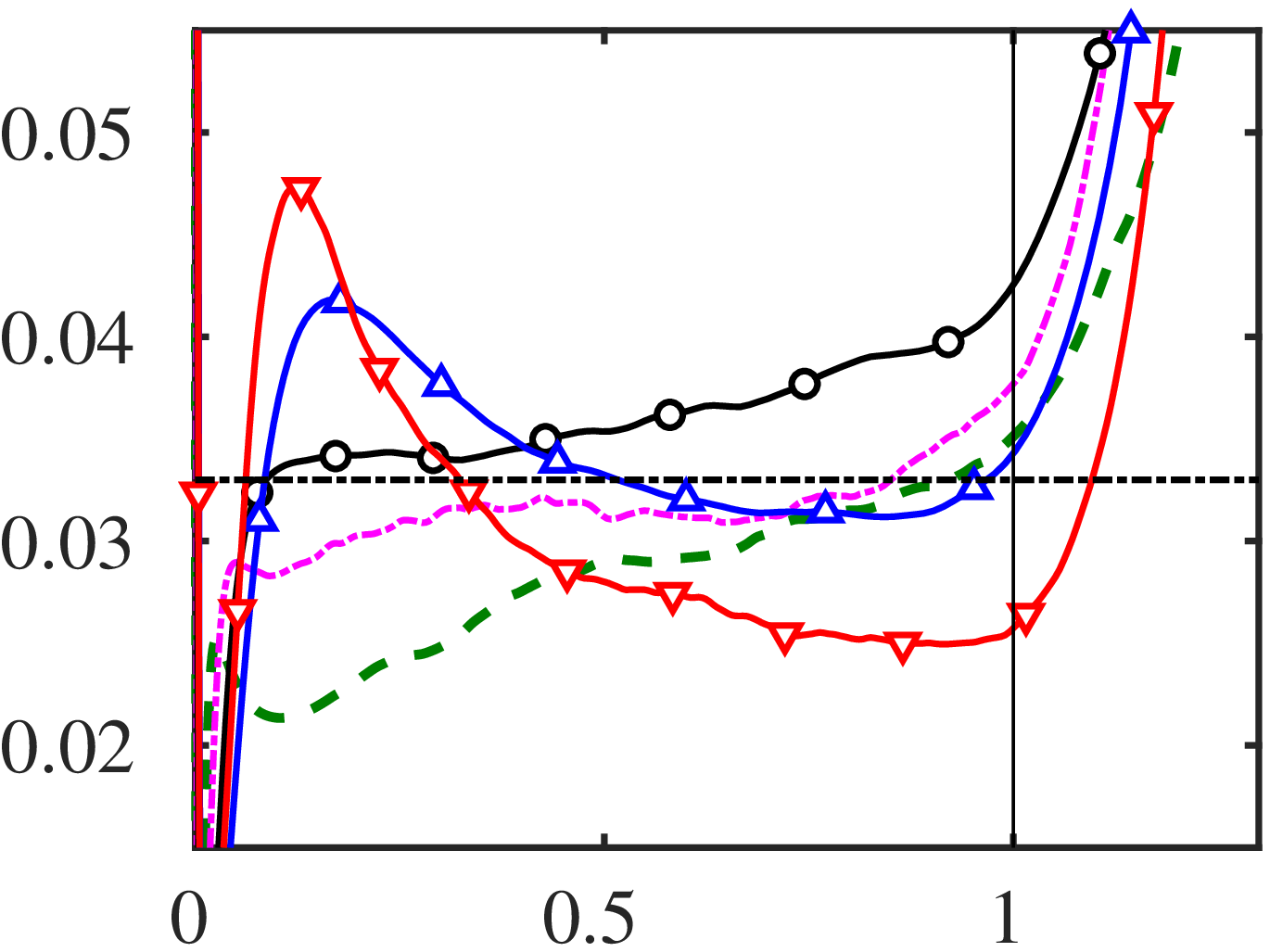}
		\centerline{$y/\delta^\ast$}
		\mylab{0.1\linewidth}{0.8\linewidth}{(d)}
	\end{minipage} 
	\caption{ 
		(a) The premultiplied Corrsin shear parameter, $S_c$, in semi-logarithmic scale. 
		The peak position corresponds to the shear thickness, $\delta^\ast$, which are shown by the dash-dotted line for each case.
		(b) The same but as a function of $y/\delta^\ast$ in the linear axis.
		\solidtridown, the logarithmic layer ($y^+ > 100 $ and $y/\delta^\ast<0.3$) for CH42 (for the clarity, only shown in (b)). 
		(c) The Corrsin length scale with respect to $\delta^\ast$. 
		(d) $(L_c/\eta) Re_\lambda^{-3/2}$. The horizontal dash-dotted line represents $0.033$ of SS-HST ~\citep{DongLozanoSekimotoJimenez2017}.     
		Lines and symbols are in table~\ref{tab:params} and the vertical solid lines in (b--d) represent $y/\delta^\ast = 1$. 
	}\label{fig:shear-rate}
\end{figure}
%
\begin{figure}
	\centering
	\begin{minipage}{3ex}
		\rotatebox[origin=c]{90}{$\bra uu \ket/U_e^2$} 
	\end{minipage} 
	\begin{minipage}{.43\linewidth}
		\includegraphics[width=1.0\linewidth,clip]{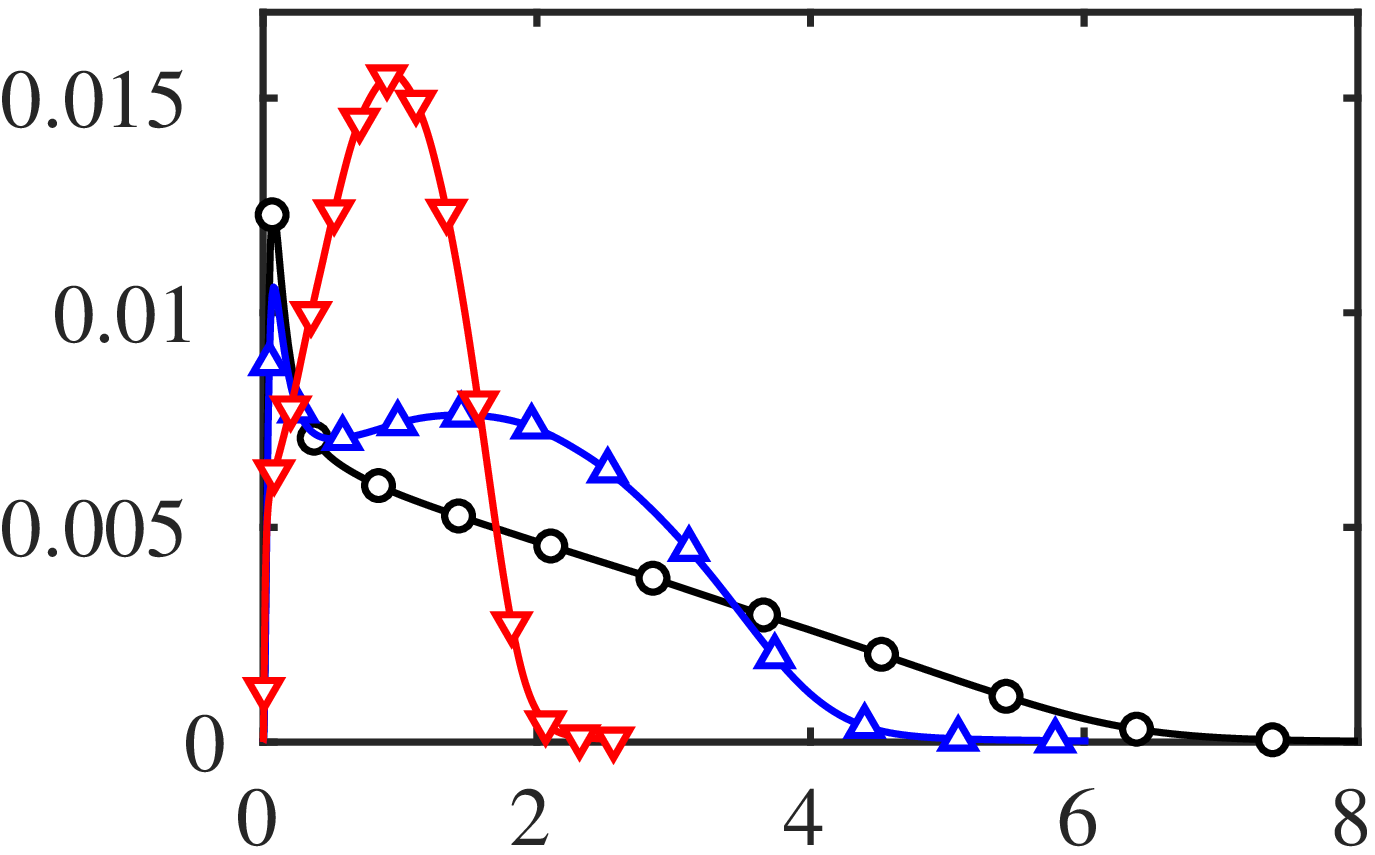}
		\centerline{$y/\delta_1$}
		\mylab{0.3\linewidth}{0.65\linewidth}{(a)}
	\end{minipage} 
	\hspace{3mm}
	\begin{minipage}{3ex}
		\rotatebox[origin=c]{90}{$\bra uu \ket/U_e^2$} 
	\end{minipage} 
	\begin{minipage}{.43\linewidth}
		\includegraphics[width=1.0\linewidth,clip]{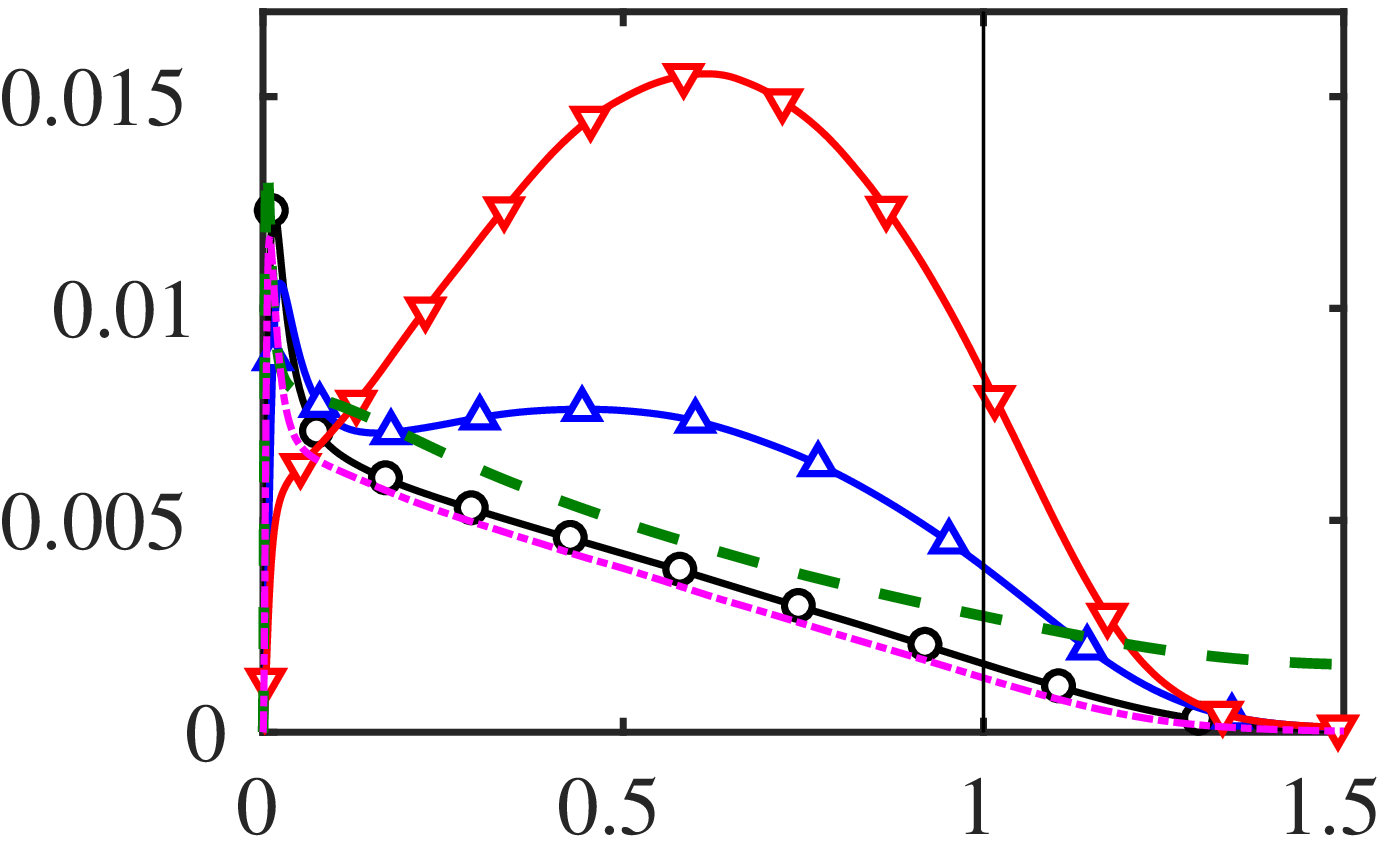}
		\centerline{$y/\delta^\ast$}
		\mylab{0.25\linewidth}{0.65\linewidth}{(b)}
	\end{minipage} 	
	\caption{ 
        Streamwise velocity fluctuation, $\bra uu \ket/U_e^2$, as a function of (a) $y/\delta_1$ and (b) $y/\delta^\ast$. 
		Lines and symbols are in table~\ref{tab:params} and the vertical solid line in (b) represents $y/\delta^\ast = 1$.
		Only two APG-TBLs and ZPG case are shown for the clarity in (a).
	}\label{fig:uu-delta1-delta3}
\end{figure}

\section{Length scale -- `the shear thickness'}\label{sec:shear-thickness}
A rigorous definition of the thickness is presented based on the nondimensional shear rate~\citep{Corrsin1958}, $S_c = \partial U/\partial y (q^2/|\varepsilon|)$,
where $q^2 = \bra u_i u_i \ket$ and $\varepsilon = - \nu \bra (\partial_j u_i) (\partial_j u_i) \ket$ is the pseudo-dissipation rate. 
Figure~\ref{fig:shear-rate}(a) shows the premultiplied form of $S_c$ as a function of $y/\delta_1$ for which the area below $(y\delta_1)S_c$ corresponds to a nondimensional velocity in the semilogarithmic plot. 
All profiles have a clear single peak, and until the peak, $S_c$ for each TBL and channel flow exhibit roughly constant values in the outer layer (see figure~\ref{fig:shear-rate}b). 
Here, the `shear thickness', $\delta^\ast$, is defined as the distance from the wall to the peak. Within this region the turbulent flow is actively sheared by the free stream. 
This definition does not rely on any arbitrary choice of threshold like $\delta_{99}$, and
it is applicable to the other configurations of inhomogeneous shear turbulence. 
In the following, we rescale the self-similar statistics of APG-TBLs using $\delta^\ast$, in comparison with the ZPG-TBLs~\citep{SilleroJimenezMoser2013,KitsiosSekimotoAtkinsonSilleroBorrellGungorJimenezSoria2017} and 
channel databases at $Re_\tau \equiv u_\tau h/\nu \approx 4200$~\citep{LozanoJimenez2014pof},
where $h$ is the channel half width. The rescaling factor $\delta^\ast/\delta_1$ is summarised in table~\ref{tab:params}.

As shown in figure~\ref{fig:shear-rate}(b), $S_c$ in the outer layer ($0.1 < y/\delta^\ast < 1$) of APG-TBLs is  $S_c \approx 7$--$9$, in good agreement with both SS-HST~\citep{SekimotoDongJimenez2016} and in the logarithmic layer of channel flow~\citep{Jimenez2013nearwall}. 
Therefore, it is considered that the region $y<\delta^\ast$ is an `active' shear-driven layer, 
and the region $y>\delta^\ast$ is `inactive' decaying turbulence.

Next, we further characterise the length scale in TBLs using $\delta^\ast$.
The Corrsin length scale, $L_c \equiv ( |\varepsilon|/(\partial_y U)^3 )^{1/2}$, 
which is shown in figure~\ref{fig:shear-rate}(c), physically represents a local characteristic large length scale, below which eddies are decoupled from the effect of the mean shear~\citep{Corrsin1958}. 
Recently, \citet{DongLozanoSekimotoJimenez2017} have shown that $L_c$ represents 
the characteristic length scale of intense Reynolds stress structures both in wall-bounded channel flow and an unbounded homogeneous shear flow. It increases linearly in the logarithmic layer of a turbulent channel flow, and keeps increasing in the wake. On the other hand, $L_c/\delta^\ast$ is roughly 0.1 in the outer layer for ZPG-TBLs and APG-TBLs for $0.3<y/\delta^\ast<1$, 
although a strong APG has a tendency of decreasing $L_c/\delta^\ast$. 
For CH42, $L_c/\delta^\ast$ increases monotonically from 0.1 to 0.2 in the range of $0.3<y/\delta^\ast<1$, 
indicating that the internal nature of channel flow creates a larger scale compared to TBLs.

The ratio of $L_c$ with respect to the Kolmogorov length,  $\eta=(\nu^3/|\varepsilon|)^{1/4}$, is shown in figure~\ref{fig:shear-rate}(d), and it represents an effective Reynolds number, $L_c/\eta \sim Re_\lambda^{3/2}$, where $Re_\lambda = q^2 (5/3\nu |\varepsilon| )^{1/2} $ is the Taylor micro-scale Reynolds number. 
In APG-TBLs, a strong inhomogeneity of scale separation is observed. 
Despite the wide range of $\beta$, however, all TBLs and turbulent channel flow 
agree roughly with that in SS-HST, $L_c/\eta \approx 0.033 Re_\lambda^{3/2}$~\citep{DongLozanoSekimotoJimenez2017}
on the average in the active outer layer.
The higher Reynolds number ZPG-TBL (BL65) shows better agreements with homogeneous shear turbulence. 
As already noted in the previous studies (e.g.,~{\cite{MacielSimensGungor2016}}), 
sweeping motions are more dominant than ejections near the wall in APG-TBLs.
The near wall peak around $y/\delta^\ast=0.1$ for APG-TBLs in figure~\ref{fig:shear-rate}(d) are consequences of the sweeps which transport the outer information of large-scale structure to near the wall, 
generating larger scale-separation near the wall. 
It is interesting that these derivation occurs around the value of homogeneous shear turbulence.
%
It should be noted that $L_c$ increases for $y>\delta^\ast$, since the mean velocity gradient decreases and the turbulence is not actively self-sustained.

\begin{figure}
	\centering
	\begin{minipage}{3ex}
		\rotatebox[origin=c]{90}{$\bra uu \ket/u^{\ast 2}$} 
	\end{minipage} 
	\begin{minipage}{.4\linewidth}
		\includegraphics[width=1.0\linewidth,clip]
		{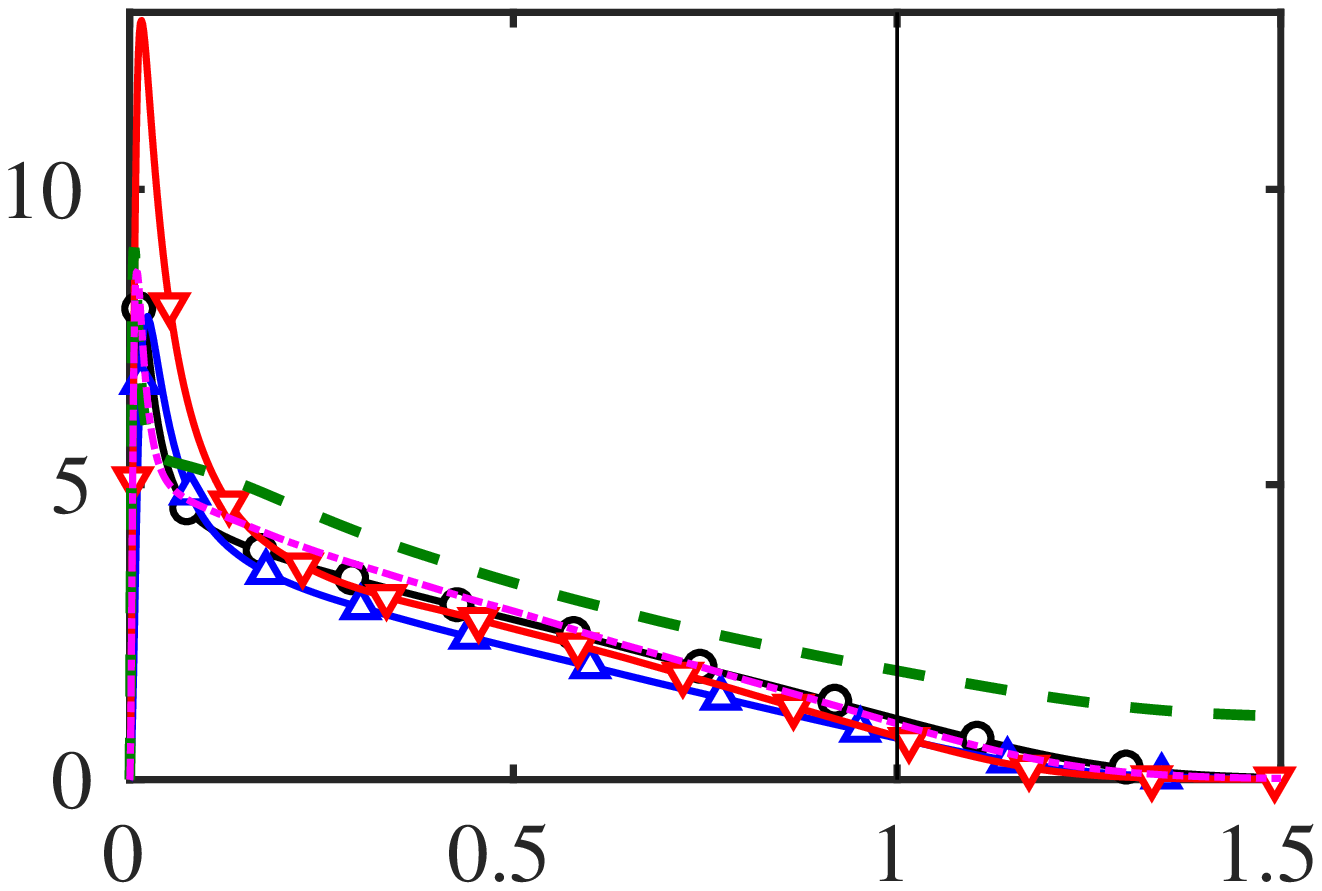}
		\centerline{$y/\delta^\ast$}
		\mylab{0.3\linewidth}{0.8\linewidth}{(a)}
	\end{minipage} 
    \hspace{1mm} 
    \begin{minipage}{3ex} 
        \rotatebox[origin=c]{90}{$ -\bra uv \ket /u^{\ast 2} $} 
    \end{minipage}
    \begin{minipage}{.42\linewidth}
        \includegraphics[width=1.0\linewidth,clip]
        {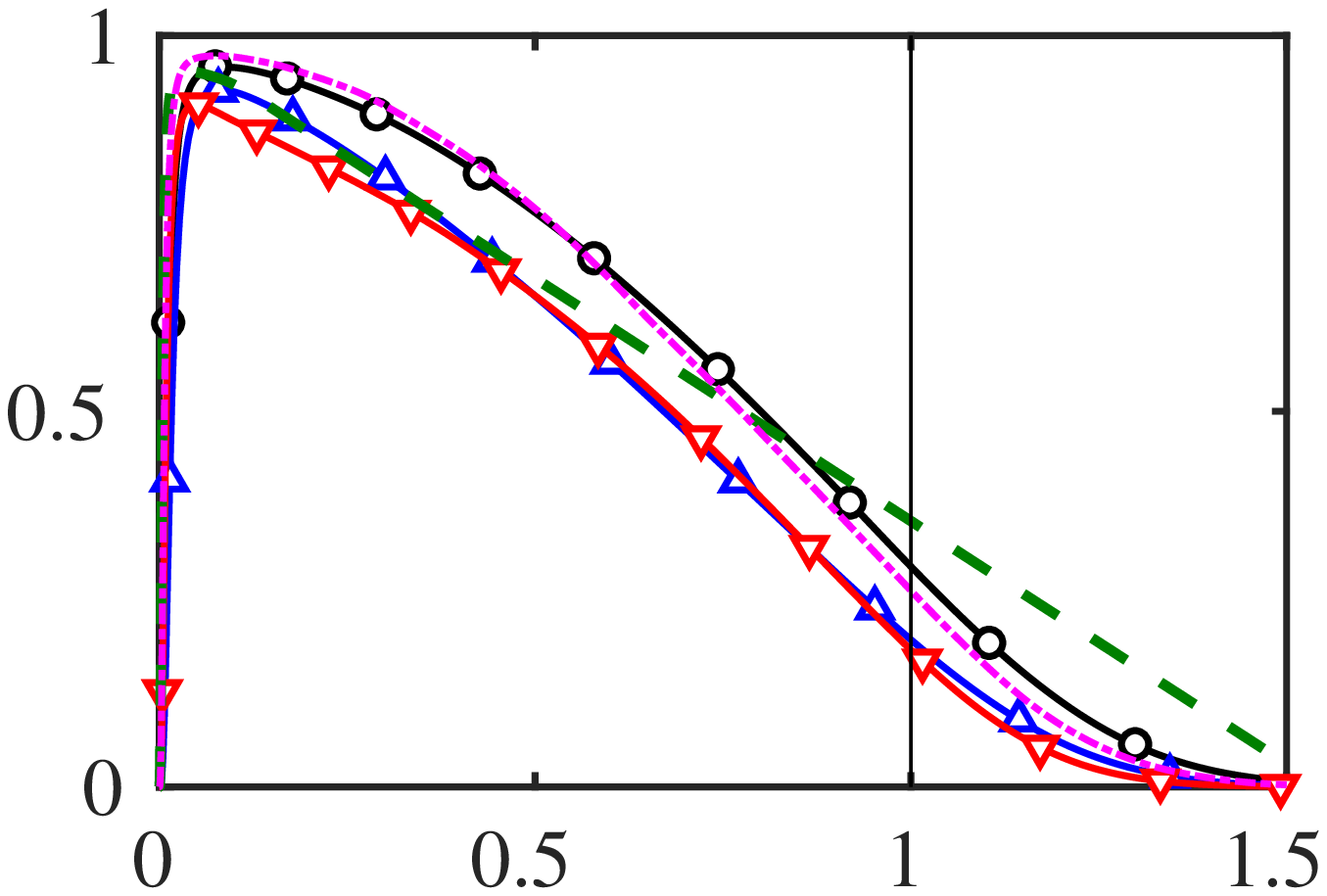}
        \centerline{$y/\delta^\ast$}
        \mylab{0.3\linewidth}{0.8\linewidth}{(b)}
    \end{minipage}  
    \\
	\begin{minipage}{3ex} 
		\rotatebox[origin=c]{90}{$ \bra vv \ket /u^{\ast 2} $} 
	\end{minipage}
	\begin{minipage}{.4\linewidth}
		\includegraphics[width=1.0\linewidth,clip]
		{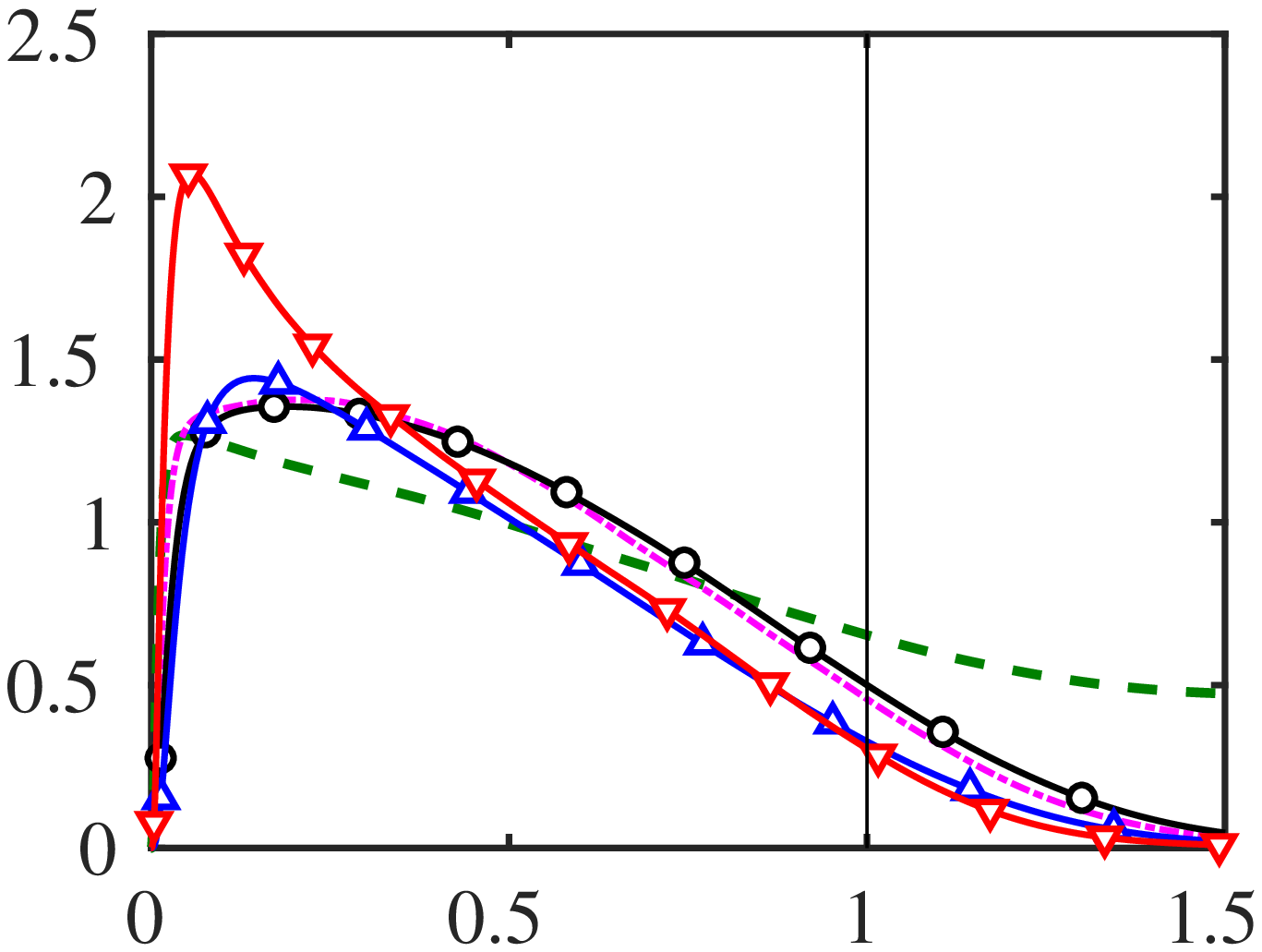}
		\centerline{$y/\delta^\ast$}
		\mylab{0.3\linewidth}{0.8\linewidth}{(c)}
	\end{minipage}
	\hspace{5mm}
	\begin{minipage}{3ex} 
    	\rotatebox[origin=c]{90}{$ \bra ww \ket /u^{\ast 2} $} 
	\end{minipage}
		\begin{minipage}{.4\linewidth}
		\includegraphics[width=1.0\linewidth,clip]
		{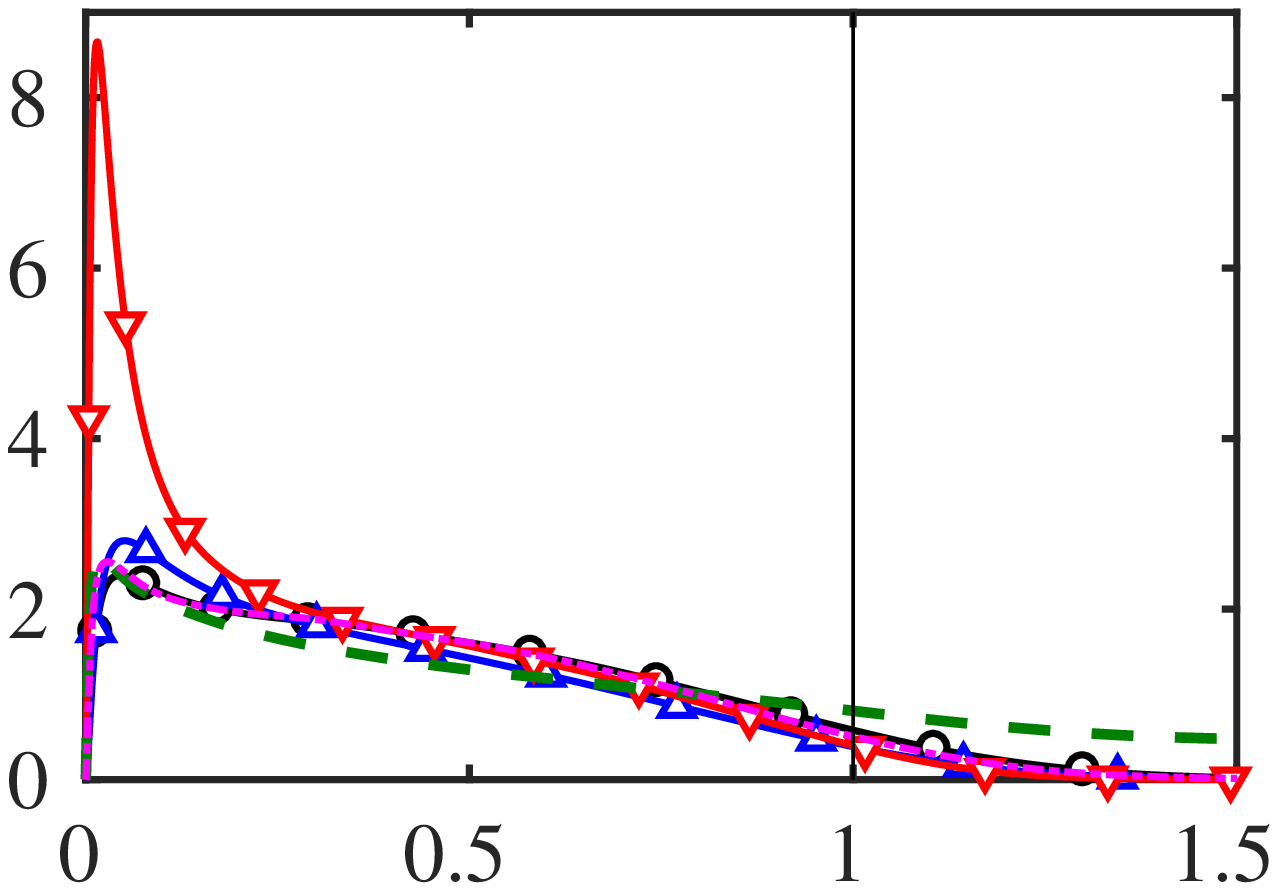}
		\centerline{$y/\delta^\ast$}
		\mylab{0.3\linewidth}{0.8\linewidth}{(d)}
	\end{minipage}  
	\caption{
		Streamwise averaged (a) $\langle uu \rangle $, 
		(b) $- \langle uv \rangle $,  
		(c) $\langle vv \rangle $, and
		(d) $\langle ww \rangle $ scaled by $u^{\ast 2}$. 
		Lines and symbols are in table~\ref{tab:params}.
	}\label{fig:turb-ustar-scaling}
\end{figure}

The streamwise-averaged velocity fluctuation, $\bra uu \ket/U_e^2$ is shown in Fig.~\ref{fig:uu-delta1-delta3}, in which the peak positions of velocity fluctuations are roughly at around $y=0.5\delta^\ast$ both for the mild and strong APG-TBLs as in figure~\ref{fig:uu-delta1-delta3}(b). 
This indicates that $\delta^\ast$ represents the characteristic length for TBLs.

\section{Velocity scales}\label{sec:scaling-vel}

The pressure velocity is re-defined using the `shear-thickness' as $u_P^\ast = \sqrt{ \delta^\ast P_e^\prime}$.
The effect of the pressure-gradient is properly measured by using the rescaled Clauser's pressure gradient parameter, $\beta^\ast =\beta (\delta^\ast/\delta_1) = (u_P^\ast/u_\tau)^2 = (Re^\ast_P/Re^\ast_\tau)^2$, where the friction Reynolds number, $Re^\ast_\tau = u_\tau \delta^\ast/\nu$ and the pressure Reynolds number, $Re^\ast_P = u^\ast_P \delta^\ast/\nu$ are also defined using $\delta^\ast$. 
These quantities are listed in table~\ref{tab:params}. 

The mixed friction-pressure velocity introduced by \cite{SkoteHenningson2002} is expressed as,
\begin{eqnarray} 
u^\ast = \sqrt{u_\tau^2 + u^{\ast 2}_P \left( \frac{y}{\delta^\ast} \right)}  
 &=& u_\tau \sqrt{1 + \beta^\ast \left( \frac{y}{\delta^\ast} \right) } \nonumber \\
 &=& u_P^\ast \sqrt{ \frac{1}{\beta^\ast} + \frac{y}{\delta^\ast}}. \label{eq:ustar}
\end{eqnarray}
The definition of $u^\ast$ is attributed to the momentum balance equation.
    The momentum equation for a simple parallel flow is given by, 
    \begin{equation}
      \frac{\mathrm{d} \bra u v \ket}{\mathrm{d} y} = - P_e^\prime + \nu \frac{\mathrm{d}^2 U}{\mathrm{d} y^2 } + F_x, \label{eq:momentum_balance_parallel}
    \end{equation}
	where $F_x$ is the forcing term to drive the flow.
	When integrated from $y=0$ to a particular position $y^\prime$ in the outer layer,	  
	the local total Reynolds stress balances as 
	\begin{equation}
	-\bra u v \ket + \nu \frac{\mathrm{d} U}{\mathrm{d} y} |_{y=y^\prime} + F_x y^\prime = u_\tau^2 + u_P^{\ast 2} \left( \frac{y^\prime}{\delta^\ast} \right) \equiv u^{\ast 2}.
	\end{equation}

%
%

For the zero-pressure-gradient case, the mean velocity profile normalised by $\nu$ and $u_\tau$ represents well-know overlap logarithmic layer between viscous sublayer and outer layer. 
The logarithmic region, however, becomes shorter with increasing $\beta^\ast$ and almost disappears for the strong APG case (see Appendix).
In the incipient separation case, the friction velocity ($u_\tau$) is not defined, and the mean velocity normalised by $u_\tau$ is undefined. While the $u^\ast$-scaling can be applied even when $u_\tau = 0$ at the verge of separation ($\beta \rightarrow \infty$) as it smoothly switches from $u_\tau$ to $u^\ast_P \sqrt{(y/\delta^\ast)}$. 


Figure~\ref{fig:turb-ustar-scaling}\rev{(a--d)} shows that $u^\ast$ scales the velocity fluctuations and the tangential Reynolds stress quite well in the outer region. It should be noted that, since $u^\ast$ is a linear function in $y$, the characteristic outer peaks of velocity fluctuations in APG-TBLs are eliminated and the profiles agree with those of ZPG-TBLs. 
Above the viscous and buffer layer, $-\bra u v \ket/ u^{\ast 2}$ is linear for CH42, 
and shows a curve to satisfy the far-field condition $\partial_y \bra u v \ket = 0$ for TBLs. 
There is, however, surprisingly good agreement in spite of the existence of the other terms in the momentum equation for TBLs. 
%

\begin{figure}
	\centering
	\begin{minipage}{3ex}
		\rotatebox[origin=l]{90}{$~~~ (y/\delta^\ast) \varepsilon^\ast,~ (y/\delta^\ast) P_k^\ast $} 
	\end{minipage}
	\begin{minipage}{.55\linewidth}
		\includegraphics[width=1.0\linewidth,clip]
		{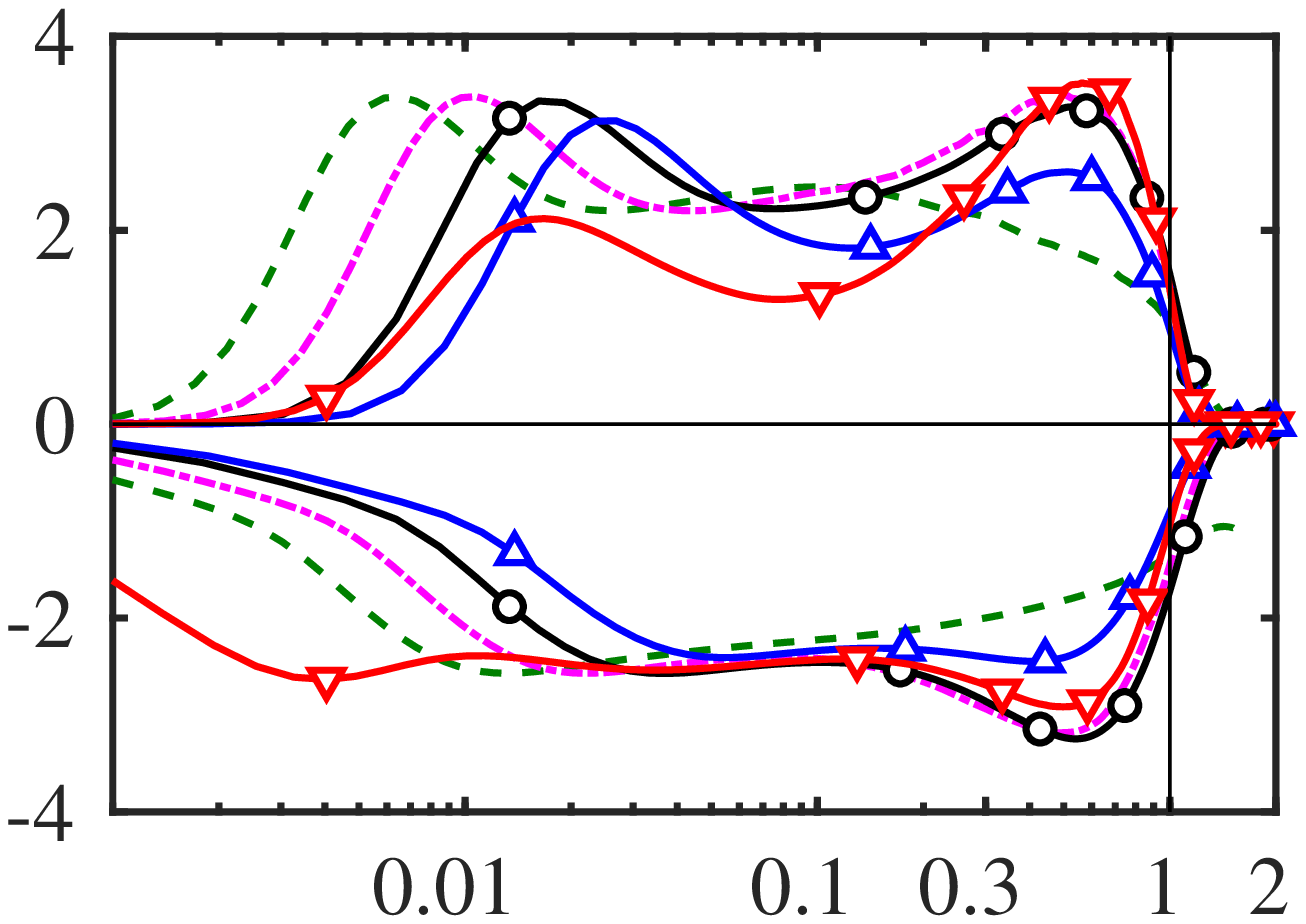}
		\centerline{$y/\delta^\ast$}
		\mylab{-0.35\linewidth}{0.8\linewidth}{$(P_k)$}
		\mylab{-0.35\linewidth}{0.3\linewidth}{$(\varepsilon)$}
	\end{minipage}
	\caption{ Premultiplied  
		kinetic energy production ($P_k$) and pseudo-dissipation terms ($\varepsilon$), 
		scaled by $u^{\ast 3}$.  
		The symbols are in table~\ref{tab:params} and the vertical solid lines in (b) represent $y/\delta^\ast = 1$.
	}\label{fig:Ke-balance}
\end{figure}

If $u^\ast$-scaling is applied to the analysis of the kinetic energy balance equation,  
the kinetic energy production, $P_k=- \bra u_i u_j \ket \partial_j U_i $, and the pseudo-dissipation rate, $\varepsilon$, are approximately balanced for $y<\delta^\ast$ as shown in figure~\ref{fig:Ke-balance}. 
On the other hand, shear-induced turbulence decays toward the irrotational flow boundary condition for $y>\delta^\ast$. 
The $u^\ast$-scaling for $P_k$ and $\varepsilon$ reveals 
the characteristic excess of the dissipation rate near the wall for strong APG-TBL 
\notyet{as in \citet{Bradshaw1967equilibriumTBL}}. 
This is possibly due to sweep motions directly impacting on the wall, since there is no near-wall high-shear buffer layer region in the strong APG.
The data of turbulent channel flow (CH40 in figure~\ref{fig:Ke-balance}) also shows a similar buffer-layer peaks in profiles of $P_k$ and $\varepsilon$ to ZPG-TBLs.


The distribution of scaled dissipation rates resemble each other within a wide range of $\beta$ as in figure~\ref{fig:Ke-balance}, indicating that the effect of adverse pressure gradients can also be eliminated from the statistics of small scale properties, i.e. vorticity. 
The scaling of vorticity magnitude, $\omega^\prime \equiv \sqrt{\omega_i \omega_i}$, in the turbulent outer region is estimated from the approximate energy balance of kinetic energy ($P_k \approx -\varepsilon$) and self-similar scaling, 
$\nu \omega^{\prime 2} \approx -\tau_{12} (\partial U/\partial y) \sim u^{\ast3}/ \delta^\ast $, where $\tau_{12}$ is the tangential Reynolds stress. 
We then have the following scaling for the vorticity magnitude,
\begin{eqnarray}
  \omega^\prime &\sim& \left( \frac{u^{\ast3}}{\nu \delta^\ast} \right)^{1/2} \nonumber \\
  &=& \left( Re_p^{\ast 2/3} \left[\frac{y}{\delta^\ast} \right] \left[\frac{u_P^\ast}{\delta^\ast}\right]^{4/3}  + Re_\tau^{\ast 2/3} \left[\frac{u_\tau}{\delta^\ast}\right]^{4/3} \right)^{3/4},\label{eq:omg-scale}
\end{eqnarray}
which leads to $\omega^\prime \sim \sqrt{Re^\ast_\tau} (u_\tau/\delta^\ast)$ for ZPG-TBL ($\beta^\ast \approx 0$) as in \cite{BorrelJimenez2016}, 
but using $\delta^\ast$ instead of $\delta_{99}$, and $\omega^\prime \sim \sqrt{Re^\ast_P} (u_P^\ast/\delta^\ast)(y/\delta^\ast)^{3/4}$ for $\beta^\ast \rightarrow \infty$.  
One of the applications of the proposed scaling is the identification of turbulent/non-turbulent interfaces using the scaled vorticity magnitude (or enstrophy) fields, which drastically decays in the inactive intermittent region given by $y>\delta^\ast$.

\section{Summary and discussions}\label{sec:discussion}
The main motivation of the present study stems from a lack of universal scaling of the statistics of adverse-pressure-gradient (APG) turbulent boundary layers (TBLs) over a wide range of $\beta$. 
Neither the classical definition of the boundary layer thickness, $\delta_{99}$, nor any thickness based on the integral of the mean velocity profile represents a characteristic length scale in TBLs, in a consistent manner.
Here, a new characteristic `shear thickness', $\delta^\ast$, is introduced, 
which divides the outer layer of TBLs into an active and an inactive one. 
The `shear thickness' is applicable to other inhomogeneous shear flows, not only wall-bounded turbulence, but also free shear flows like jets and mixing layers. 
There are well-known characteristic peaks of the velocity fluctuations in APG-TBLs, and their positions coincide when they are scaled by $y/\delta^\ast$ (see Fig.~\ref{fig:uu-delta1-delta3}), implying that the present $\delta^\ast$ successfully indicates a characteristic outer-length scale over a wide range of APGs.
The active outer layer $y<\delta^\ast$ in TBLs represents an approximately constant Corrsin shear-rate parameter, in which the local mean shear contributes to the energy production and organisation of large-scale motions, while, the inactive outer-outer layer $y>\delta^\ast$ is attributed to be a region where shear-induced turbulence is decaying and interacting with the free-stream.
The Corrsin length scale, which represents the local large length scale of shear-induced turbulence, is approximately $0.1 \delta^\ast$ in the outer layer of TBLs over a wide range of $\beta$ (see figure~\ref{fig:shear-rate}c). This also agrees well with the top of the logarithmic layer of turbulent channel flow ($y/\delta^\ast \approx 0.3 $). 
The local scale separation $(L_c/\eta)$ normalised by $Re_\lambda^{3/2}$ shows inhomogeneity in the outer layer of APG-TBLs. However, there is still reasonable agreement with both the outer layer of turbulent channel flow ($0.3 < y/\delta^\ast < 1.0$) and statistically-stationary homogeneous shear turbulence ~\citep{DongLozanoSekimotoJimenez2017}. This indicates that the largest scale motions, which determine the outer length scale, are common in all turbulent shear flows.

The characteristic peaks of the velocity fluctuations in the outer layer of APG-TBLs are disappeared by using the mixed friction-pressure scale,
$u^\ast = \sqrt{u_\tau^2 + u_P^{\ast 2}(y/\delta^\ast) }$,
which is found to scale
velocity fluctuations and Reynolds stresses in the outer-layer, $0.3<y/\delta^\ast$, for all TBLs (see figure~\ref{fig:turb-ustar-scaling}), collapsing to the zero-pressure-gradient cases. 
The $u^\ast$-scaling is used to characterise the kinetic energy equation of TBLs and channel flows.
The kinetic energy production and dissipation rate are locally balanced and have similar distributions for all TBLs and channel flow except for the near-wall excess of the dissipation rate in the strong APG-TBL (see figure~\ref{fig:Ke-balance}).  
The general form of the `star-unit' scaling for the vorticity magnitude 
in TBLs is also derived. 

It has been shown that the APG effect is eliminated by the proper friction-pressure mixed scaling in the outer layer, implying that there is a common mechanism in shear-induced turbulent flows. 
%
\notyet{There is an additional history effect in normal TBL flows, 
whose effects can significantly impact on the statistics, and need to be clarified based on the present scaling as a future work.}
We believe that the present scaling provides new physical insights applicable to the development of turbulence models.

\section*{Acknowledgements}
 The authors would like to acknowledge the research funding from the Australian
 Research Council, and the computational resources provided by the National Computational Infrastructure through a NCMAS grant and PRACE. 
 The authors would like to acknowledge Prof. Javier Jim\'enez for his fruitful discussions and insights on this topic, and Dr. Lozano-Dur\'an for his support to plot the database of turbulent channel flows.
 Atsushi Sekimoto would like to acknowledge the supports of JSPS KAKENHI Grant-in-Aid for Early-Career Scientists (T18K136890) and International Aircraft Development Fund.
 Callum Atkinson acknowledges the support of an Australian Research Council Discovery Early Career Researcher 
 Award Fellowship. Julio Soria gratefully acknowledges the support of an Australian
 Research Council Discovery Outstanding Researcher Award fellowship.
\\
\appendix
\section{Scaling for the mean velocity}
\begin{figure}
	\begin{minipage}{3ex}
		\rotatebox[origin=c]{90}{$ U/U_e $} 
  	\end{minipage}
	\begin{minipage}{.46\linewidth}
		\includegraphics[width=1.0\linewidth,clip]
		{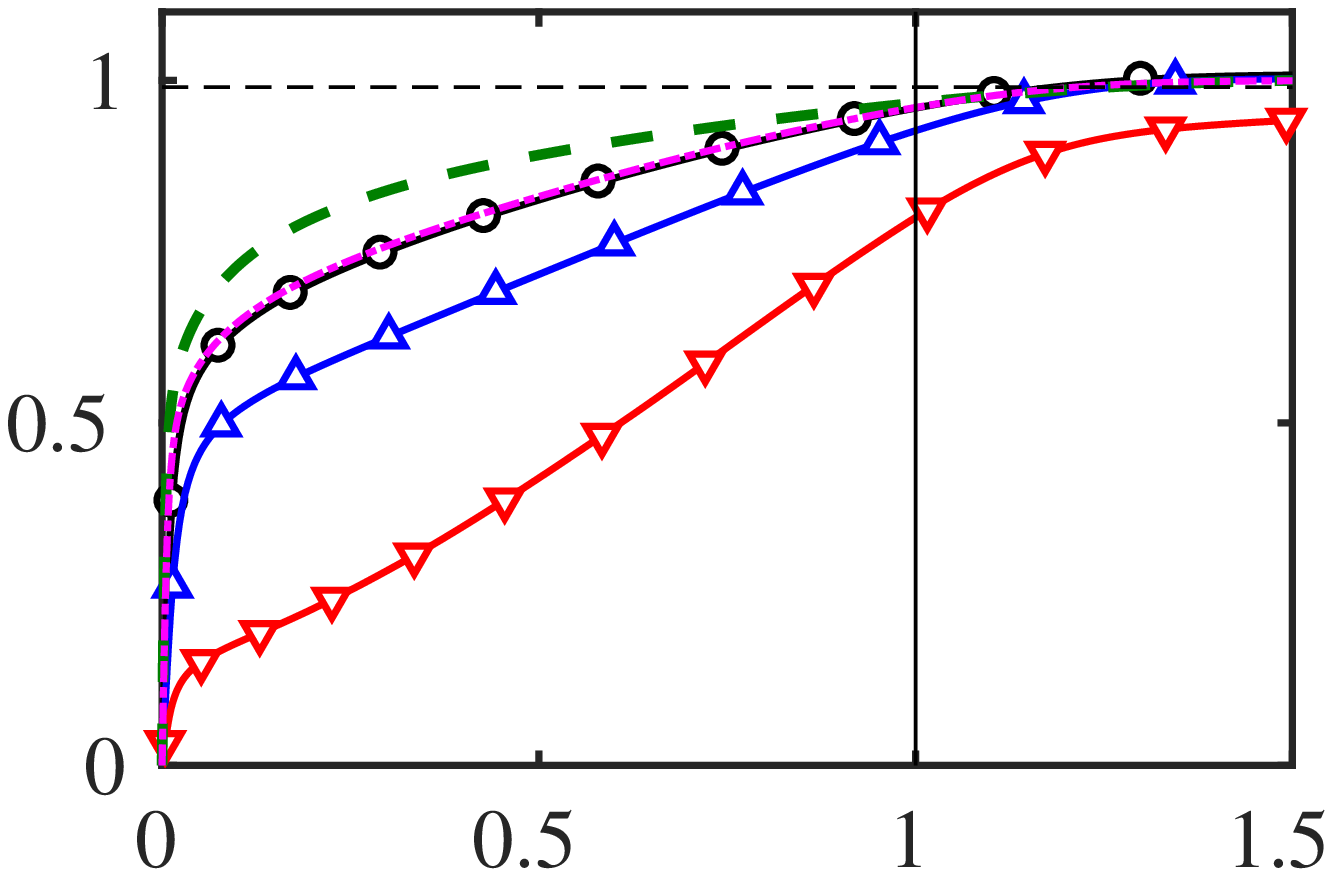}
		\centerline{$y/\delta^\ast$}
		\mylab{0.3\linewidth}{0.35\linewidth}{(a)}
	\end{minipage}
	\hspace{1mm} 
	\begin{minipage}{3ex}
		\rotatebox[origin=c]{90}{$ (U_e - U)/u_0 $} 
	\end{minipage}
	\begin{minipage}{.4\linewidth}
		\includegraphics[width=1.0\linewidth,clip]
		{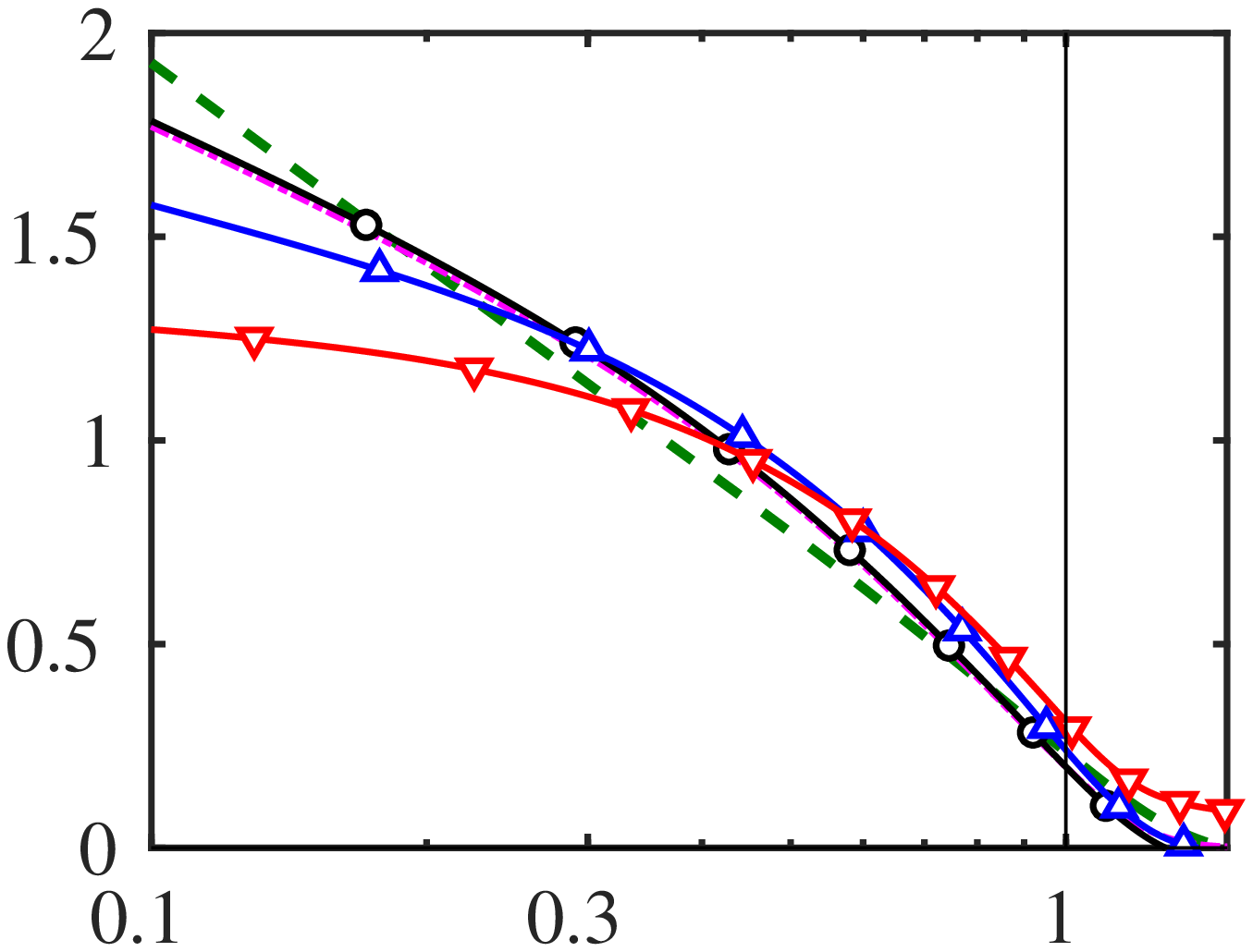}
		\centerline{$y/\delta^\ast$}
		\mylab{0.35\linewidth}{0.85\linewidth}{(b)}
	\end{minipage} 
 \\
	\hspace{1mm} 
    \begin{minipage}{3ex}
        \rotatebox[origin=c]{90}{$ U/(\alpha u^\ast)$} 
    \end{minipage}
	\hspace{1mm} 
    \begin{minipage}{.43\linewidth}
        \includegraphics[width=1.0\linewidth,clip]
        {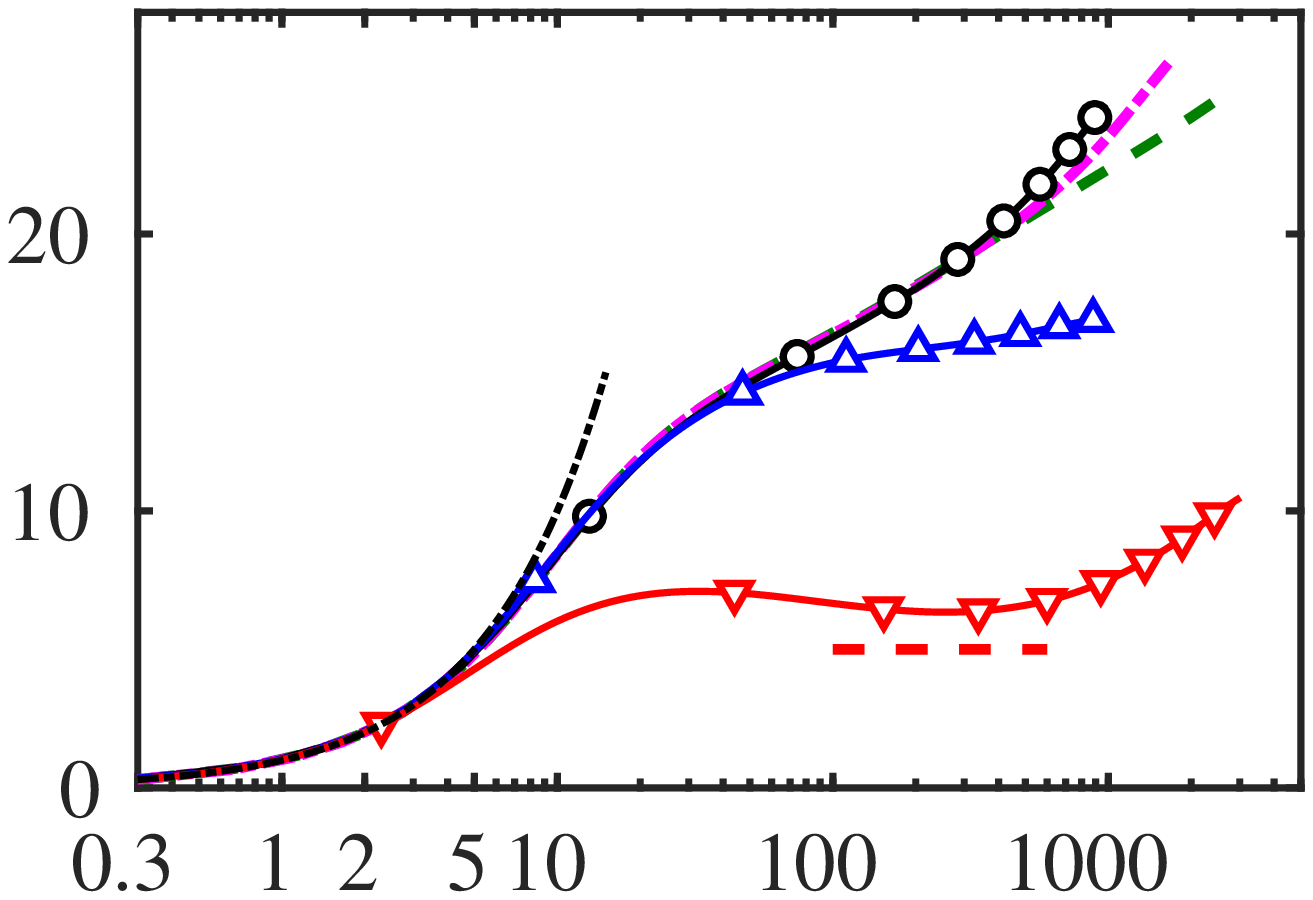}
        \centerline{$y (\alpha u^\ast/\nu)$}
        \mylab{-0.3\linewidth}{0.8\linewidth}{(c)}
    \end{minipage}
	\hspace{3mm} 
	\begin{minipage}{3ex}
        \rotatebox[origin=c]{90}{$ (U_e-U)\delta^\ast/(\alpha u^\ast\delta_1) $} 
    \end{minipage}
    \begin{minipage}{.42\linewidth}
        \includegraphics[width=1.0\linewidth,clip]
        {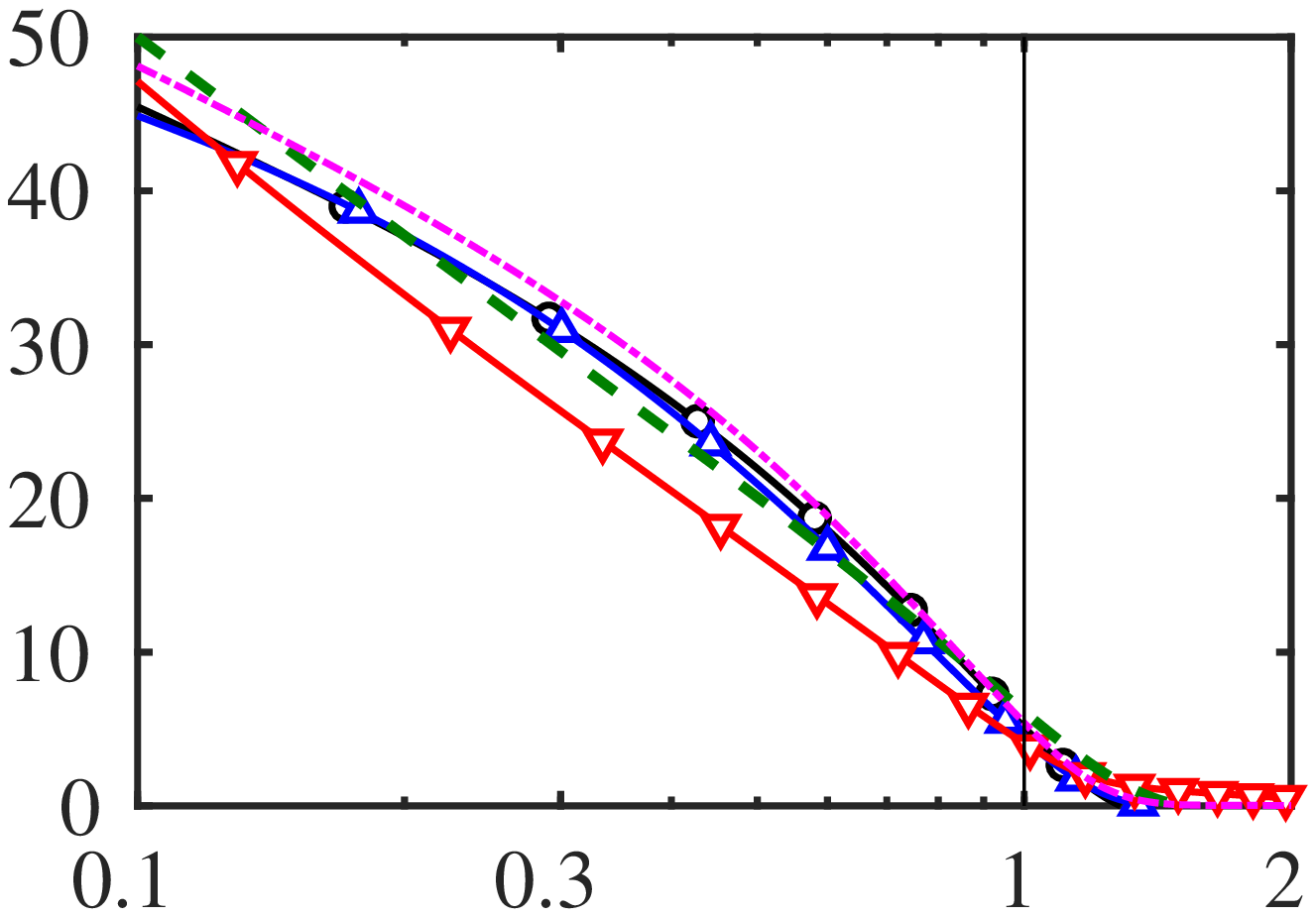}
        \centerline{$y/\delta^\ast$}
        \mylab{0.3\linewidth}{0.8\linewidth}{(d)}
    \end{minipage} 
	\caption{ Streamwise averaged (a) mean velocity scaled by $U_e$, (b) mean velocity deficit scaled by $u_0 = U_e(\delta_1/\delta^\ast)$,  The horizontal dashed line in (a) represents $0.99U_e$. 
	\rev{(c) Streamwise averaged mean velocity scaled by $\alpha u^\ast $ and $\nu/(\alpha u^\ast)$. 
	The black dash-dotted line represents the viscous limit, $U/(\alpha u^\ast) = y (\alpha u^\ast/\nu)$, and the horizontal red dashed line represents $U/(\alpha u^\ast) = 2/\kappa$. (d) The velocity deficit $(U_e - U)$ scaled by $\alpha u^\ast (\delta_1/\delta^\ast)$ as functions of $y/\delta^\ast$.}
	Lines and symbols are in table~\ref{tab:params} \notyet{and the vertical solid lines in (a, b, d) represent $y/\delta^\ast = 1$}.  
	}\label{fig:mean-velocity}
\end{figure}

\rev{
In this section, in case that some readers might be interested in scaling the mean velocity using the present outer length $\delta^\ast$ and the corresponding velocity scales. }
Figure~\ref{fig:mean-velocity}(a,b) shows the mean velocity profile and the velocity deficit scaled by $U_e$ and $u_0=U_e\delta_1/\delta^\ast$, respectively. $u_0$ is a bulk velocity, i.e. the flow rate divided by $\delta^\ast$, and is analogous to the Zagarola-Smits scaling, $U_{ZS} = U_e\delta_1/\delta_{99}$~\citep{ZagarolaSmitsFEDSM1998}, but using $\delta^\ast$ rather than $\delta_{99}$.  It appears to be the appropriate velocity scale for the velocity deficit. 
\notyet{The reason why this Zagarola-Smits scaling works is explained in \citet{WeiMaciel2018} for zero-pressure-gradient case.} 
\rev{\citet{LogdbergAngeleAlfredsson2008} also shows the good collapse for 
the mean velocities with a wide range of pressure gradient including separated flow.
}
  
\rev{
Next, we show a couple of trials of the present $u^\ast$-scaling for the mean velocity profile using the present database. }
As in \citet{KitsiosSekimotoAtkinsonSilleroBorrellGungorJimenezSoria2017}, the mean velocity in the vicinity of the wall can be expressed by 
$ U = (1/2\nu) u^{\ast 2}_P (y^2/\delta^\ast) + (u_\tau^2/\nu) y $, so that $ U/u^\ast = \alpha^2 (y u^\ast/\nu) $, where $\alpha^2=(1 + u_\tau^2/u^{\ast 2})/2$.
The limits of the mean velocity profile in the vicinity of the wall are $\alpha^2 = 1$ for the flow with ZPG ($\beta=0$) and $\alpha^2 = 1/2 $ for the flow at the verge of separation ($ \beta \rightarrow \infty $).
\rev{This idea also covers the viscous layer of separated flow by using a negative value of $\alpha$}.
Figure~\ref{fig:mean-velocity}(c) shows the mean velocity profile normalised by $\alpha u^\ast$, 
where the distance from the wall is normalised by $l_\nu^\ast \equiv \nu/(\alpha u^\ast)$, i.e.
\begin{equation}
\frac{y}{l_\nu^\ast} = \left( \frac{y}{\delta^\ast} \right) 
\sqrt{Re_\tau^{\ast 2} + \frac{Re_P^{\ast 2}}{2} \left( \frac{y}{\delta^\ast} \right) }. 
\end{equation}
At incipient separation ($\beta \rightarrow \infty$), the so-called square-root profile~\citep{Stratford1959}
is represented as a constant $U/(\alpha u^\ast) = 2/ \kappa$, 
where we used a typical value of the K\'arm\'an constant, $\kappa \approx 0.4$ (see the horizontal dashed line in figure~\ref{fig:mean-velocity}(c)). 
\notyet{The scaled mean velocity and scaled coordinate enable a continuous plot for all APG-TBLs    
	as in Fig.~\ref{fig:mean-velocity}(c) within the range of $U/(\alpha u^\ast) < 30$. }
	The inertial term (history effect), which is ignored in eq.~(\ref{eq:momentum_balance_parallel}) could be an important factor to estimate the mean velocity profile. In this study, however, we leave such history effect on the mean velocity as a future work, and further investigate the universal scaling of turbulent statistics using $u^\ast$.
%
\rev{Figure~\ref{fig:mean-velocity}(d) is the mean velocity deficit scaled by $\alpha u^\ast (\delta_1/\delta^\ast)$. The mean velocity in the viscous- and outer-layer appears to be scaled as in Fig.~\ref{fig:mean-velocity}(c) and (d), which might imply an overlap region for the mean velocity in APG-TBLs.}   
\notyet{There were some proposals for the modification of logarithmic mean velocity profile using the pressure gradient parameter~\cite{SkoteHenningson2002}, 
however, our experiments to obtain an generalised wall function over a wide range of APG were not successful. The Reynolds number may not be high enough to obtain the overlap region for all APG-TBL.} 
	Finding a generalised wall function over a wide range of APG using $u^\ast$ and the corresponding length scale, $\nu/u^\ast$, is an ambitious problem, however, our experiments attempting to find an appropriate generalized wall function of the mean velocity have not been successful. Therefore, the question of how the mean velocity is determined by APG is still an open one.

\bibliography{database_tbl}

\begin{thebibliography}{30}%
\makeatletter
\providecommand \@ifxundefined [1]{%
 \@ifx{#1\undefined}
}%
\providecommand \@ifnum [1]{%
 \ifnum #1\expandafter \@firstoftwo
 \else \expandafter \@secondoftwo
 \fi
}%
\providecommand \@ifx [1]{%
 \ifx #1\expandafter \@firstoftwo
 \else \expandafter \@secondoftwo
 \fi
}%
\providecommand \natexlab [1]{#1}%
\providecommand \enquote  [1]{``#1''}%
\providecommand \bibnamefont  [1]{#1}%
\providecommand \bibfnamefont [1]{#1}%
\providecommand \citenamefont [1]{#1}%
\providecommand \href@noop [0]{\@secondoftwo}%
\providecommand \href [0]{\begingroup \@sanitize@url \@href}%
\providecommand \@href[1]{\@@startlink{#1}\@@href}%
\providecommand \@@href[1]{\endgroup#1\@@endlink}%
\providecommand \@sanitize@url [0]{\catcode `\\12\catcode `\$12\catcode
  `\&12\catcode `\#12\catcode `\^12\catcode `\_12\catcode `\%12\relax}%
\providecommand \@@startlink[1]{}%
\providecommand \@@endlink[0]{}%
\providecommand \url  [0]{\begingroup\@sanitize@url \@url }%
\providecommand \@url [1]{\endgroup\@href {#1}{\urlprefix }}%
\providecommand \urlprefix  [0]{URL }%
\providecommand \Eprint [0]{\href }%
\providecommand \doibase [0]{https://doi.org/}%
\providecommand \selectlanguage [0]{\@gobble}%
\providecommand \bibinfo  [0]{\@secondoftwo}%
\providecommand \bibfield  [0]{\@secondoftwo}%
\providecommand \translation [1]{[#1]}%
\providecommand \BibitemOpen [0]{}%
\providecommand \bibitemStop [0]{}%
\providecommand \bibitemNoStop [0]{.\EOS\space}%
\providecommand \EOS [0]{\spacefactor3000\relax}%
\providecommand \BibitemShut  [1]{\csname bibitem#1\endcsname}%
\let\auto@bib@innerbib\@empty
\bibitem [{\citenamefont {Townsend}(1976)}]{Townsend1976book}%
  \BibitemOpen
  \bibfield  {author} {\bibinfo {author} {\bibfnamefont {A.~A.}\ \bibnamefont
  {Townsend}},\ }\href@noop {} {\emph {\bibinfo {title} {The Structure of
  Turbulent Shear Flows, 2nd ed.}}}\ (\bibinfo  {publisher} {Cambridge U.
  Press},\ \bibinfo {year} {1976})\BibitemShut {NoStop}%
\bibitem [{\citenamefont {Townsend}(1960)}]{Townsend1960}%
  \BibitemOpen
  \bibfield  {author} {\bibinfo {author} {\bibfnamefont {A.~A.}\ \bibnamefont
  {Townsend}},\ }\bibfield  {title} {\bibinfo {title} {The development of
  turbulent boundary layers with negligible wall stress},\ }\href@noop {}
  {\bibfield  {journal} {\bibinfo  {journal} {J. Fluid Mech.}\ }\textbf
  {\bibinfo {volume} {8}},\ \bibinfo {pages} {143} (\bibinfo {year}
  {1960})}\BibitemShut {NoStop}%
\bibitem [{\citenamefont {Mellor}\ and\ \citenamefont
  {Gibson}(1966)}]{MellorGibson1966}%
  \BibitemOpen
  \bibfield  {author} {\bibinfo {author} {\bibfnamefont {G.~L.}\ \bibnamefont
  {Mellor}}\ and\ \bibinfo {author} {\bibfnamefont {D.~M.}\ \bibnamefont
  {Gibson}},\ }\bibfield  {title} {\bibinfo {title} {Equilibrium turbulent
  boundary layers},\ }\href@noop {} {\bibfield  {journal} {\bibinfo  {journal}
  {J. Fluid Mech.}\ }\textbf {\bibinfo {volume} {24}},\ \bibinfo {pages} {225}
  (\bibinfo {year} {1966})}\BibitemShut {NoStop}%
\bibitem [{\citenamefont {Skote}\ \emph {et~al.}(1998)\citenamefont {Skote},
  \citenamefont {Henningson},\ and\ \citenamefont {Henkes}}]{Skote1998}%
  \BibitemOpen
  \bibfield  {author} {\bibinfo {author} {\bibfnamefont {M.}~\bibnamefont
  {Skote}}, \bibinfo {author} {\bibfnamefont {D.}~\bibnamefont {Henningson}},\
  and\ \bibinfo {author} {\bibfnamefont {R.}~\bibnamefont {Henkes}},\
  }\bibfield  {title} {\bibinfo {title} {Direct numerical simulation of
  self-similar turbulent boundary layers in adverse pressure gradients},\
  }\href {https://doi.org/10.1023/A:1009934906108} {\bibfield  {journal}
  {\bibinfo  {journal} {Flow, Turbulence and Combustion}\ }\textbf {\bibinfo
  {volume} {60}},\ \bibinfo {pages} {47} (\bibinfo {year} {1998})}\BibitemShut
  {NoStop}%
\bibitem [{\citenamefont {Kitsios}\ \emph {et~al.}(2016)\citenamefont
  {Kitsios}, \citenamefont {Atkinson}, \citenamefont {Sillero}, \citenamefont
  {Borrell}, \citenamefont {Gungor}, \citenamefont {Jim\'enez},\ and\
  \citenamefont {Soria}}]{KitsiosAtkinsonSilleroBorrellGungorJimenezSoria2016}%
  \BibitemOpen
  \bibfield  {author} {\bibinfo {author} {\bibfnamefont {V.}~\bibnamefont
  {Kitsios}}, \bibinfo {author} {\bibfnamefont {C.}~\bibnamefont {Atkinson}},
  \bibinfo {author} {\bibfnamefont {J.}~\bibnamefont {Sillero}}, \bibinfo
  {author} {\bibfnamefont {G.}~\bibnamefont {Borrell}}, \bibinfo {author}
  {\bibfnamefont {A.}~\bibnamefont {Gungor}}, \bibinfo {author} {\bibfnamefont
  {J.}~\bibnamefont {Jim\'enez}},\ and\ \bibinfo {author} {\bibfnamefont
  {J.}~\bibnamefont {Soria}},\ }\bibfield  {title} {\bibinfo {title} {Direct
  numerical simulation of a self-similar adverse pressure gradient turbulent
  boundary layer},\ }\href@noop {} {\bibfield  {journal} {\bibinfo  {journal}
  {Int. J. Heat Fluid Flow}\ }\textbf {\bibinfo {volume} {61}},\ \bibinfo
  {pages} {129} (\bibinfo {year} {2016})}\BibitemShut {NoStop}%
\bibitem [{\citenamefont {Kitsios}\ \emph {et~al.}(2017)\citenamefont
  {Kitsios}, \citenamefont {Sekimoto}, \citenamefont {Atkinson}, \citenamefont
  {Sillero}, \citenamefont {Borrell}, \citenamefont {Gungor}, \citenamefont
  {Jim\'enez},\ and\ \citenamefont
  {Soria}}]{KitsiosSekimotoAtkinsonSilleroBorrellGungorJimenezSoria2017}%
  \BibitemOpen
  \bibfield  {author} {\bibinfo {author} {\bibfnamefont {V.}~\bibnamefont
  {Kitsios}}, \bibinfo {author} {\bibfnamefont {A.}~\bibnamefont {Sekimoto}},
  \bibinfo {author} {\bibfnamefont {C.}~\bibnamefont {Atkinson}}, \bibinfo
  {author} {\bibfnamefont {J.}~\bibnamefont {Sillero}}, \bibinfo {author}
  {\bibfnamefont {G.}~\bibnamefont {Borrell}}, \bibinfo {author} {\bibfnamefont
  {A.~G.}\ \bibnamefont {Gungor}}, \bibinfo {author} {\bibfnamefont
  {J.}~\bibnamefont {Jim\'enez}},\ and\ \bibinfo {author} {\bibfnamefont
  {J.}~\bibnamefont {Soria}},\ }\bibfield  {title} {\bibinfo {title} {Direct
  numerical simulation of a self-similar adverse pressure gradient turbulent
  boundary layer at the verge of separation},\ }\href@noop {} {\bibfield
  {journal} {\bibinfo  {journal} {J. Fluid Mech.}\ }\textbf {\bibinfo {volume}
  {829}},\ \bibinfo {pages} {392} (\bibinfo {year} {2017})}\BibitemShut
  {NoStop}%
\bibitem [{\citenamefont {Lighthill}(1963)}]{Lighthill1963book}%
  \BibitemOpen
  \bibfield  {author} {\bibinfo {author} {\bibfnamefont {M.~J.}\ \bibnamefont
  {Lighthill}},\ }\href@noop {} {\emph {\bibinfo {title} {Introduction.
  boundary layer theory. In {\it Laminar Boundary Layers}}}}\ (\bibinfo
  {publisher} {London: Oxford Univ. Press},\ \bibinfo {year}
  {1963})\BibitemShut {NoStop}%
\bibitem [{\citenamefont {Spalart}\ and\ \citenamefont
  {Watmuff}(1993)}]{SpalartWatmuff1993}%
  \BibitemOpen
  \bibfield  {author} {\bibinfo {author} {\bibfnamefont {P.~R.}\ \bibnamefont
  {Spalart}}\ and\ \bibinfo {author} {\bibfnamefont {J.~H.}\ \bibnamefont
  {Watmuff}},\ }\bibfield  {title} {\bibinfo {title} {Experimental and
  numerical study of a turbulent boundary layer with pressure gradients},\
  }\href@noop {} {\bibfield  {journal} {\bibinfo  {journal} {J. Fluid Mech.}\
  }\textbf {\bibinfo {volume} {249}},\ \bibinfo {pages} {337} (\bibinfo {year}
  {1993})}\BibitemShut {NoStop}%
\bibitem [{\citenamefont {Vinuesa}\ \emph {et~al.}(2016)\citenamefont
  {Vinuesa}, \citenamefont {Bobke}, \citenamefont {{\"O}rl{\"u}},\ and\
  \citenamefont {Schlatter}}]{VinuesaBobkeOrluSchlatter2016}%
  \BibitemOpen
  \bibfield  {author} {\bibinfo {author} {\bibfnamefont {R.}~\bibnamefont
  {Vinuesa}}, \bibinfo {author} {\bibfnamefont {A.}~\bibnamefont {Bobke}},
  \bibinfo {author} {\bibfnamefont {R.}~\bibnamefont {{\"O}rl{\"u}}},\ and\
  \bibinfo {author} {\bibfnamefont {P.}~\bibnamefont {Schlatter}},\ }\bibfield
  {title} {\bibinfo {title} {On determining characteristic length scales in
  pressure-gradient turbulent boundary layers},\ }\href@noop {} {\bibfield
  {journal} {\bibinfo  {journal} {Phys. Fluids}\ }\textbf {\bibinfo {volume}
  {28}},\ \bibinfo {pages} {055101} (\bibinfo {year} {2016})}\BibitemShut
  {NoStop}%
\bibitem [{\citenamefont {Alfredsson}\ and\ \citenamefont
  {{\"O}rl{\"u}}(2010)}]{AlfredssonOrlu2010}%
  \BibitemOpen
  \bibfield  {author} {\bibinfo {author} {\bibfnamefont {P.~H.}\ \bibnamefont
  {Alfredsson}}\ and\ \bibinfo {author} {\bibfnamefont {R.}~\bibnamefont
  {{\"O}rl{\"u}}},\ }\bibfield  {title} {\bibinfo {title} {The diagnostic plot
  -- a litmus test for wall bounded turbulence data},\ }\href@noop {}
  {\bibfield  {journal} {\bibinfo  {journal} {Eur. J. Mech., B/Fluids}\
  }\textbf {\bibinfo {volume} {29}},\ \bibinfo {pages} {403 } (\bibinfo {year}
  {2010})}\BibitemShut {NoStop}%
\bibitem [{\citenamefont {Gungor}\ \emph {et~al.}(2016)\citenamefont {Gungor},
  \citenamefont {Maciel}, \citenamefont {Simens},\ and\ \citenamefont
  {Soria}}]{GungorMacielSimensSoria2016}%
  \BibitemOpen
  \bibfield  {author} {\bibinfo {author} {\bibfnamefont {A.}~\bibnamefont
  {Gungor}}, \bibinfo {author} {\bibfnamefont {Y.}~\bibnamefont {Maciel}},
  \bibinfo {author} {\bibfnamefont {M.}~\bibnamefont {Simens}},\ and\ \bibinfo
  {author} {\bibfnamefont {J.}~\bibnamefont {Soria}},\ }\bibfield  {title}
  {\bibinfo {title} {Scaling and statistics of large-defect adverse pressure
  gradient turbulent boundary layers},\ }\href@noop {} {\bibfield  {journal}
  {\bibinfo  {journal} {Int. J. Heat Fluid Flow}\ }\textbf {\bibinfo {volume}
  {59}},\ \bibinfo {pages} {109} (\bibinfo {year} {2016})}\BibitemShut
  {NoStop}%
\bibitem [{\citenamefont {Zagarola}\ and\ \citenamefont
  {Smits}(1998)}]{ZagarolaSmitsFEDSM1998}%
  \BibitemOpen
  \bibfield  {author} {\bibinfo {author} {\bibfnamefont {M.}~\bibnamefont
  {Zagarola}}\ and\ \bibinfo {author} {\bibfnamefont {A.}~\bibnamefont
  {Smits}},\ }\bibfield  {title} {\bibinfo {title} {A new mean velocity scaling
  for turbulent boundary layers},\ }in\ \href@noop {} {\emph {\bibinfo
  {booktitle} {Proc. of FEDSM'98, Washington DC}}}\ (\bibinfo {year}
  {1998})\BibitemShut {NoStop}%
\bibitem [{\citenamefont {Skote}\ and\ \citenamefont
  {Henningson}(2002)}]{SkoteHenningson2002}%
  \BibitemOpen
  \bibfield  {author} {\bibinfo {author} {\bibfnamefont {M.}~\bibnamefont
  {Skote}}\ and\ \bibinfo {author} {\bibfnamefont {D.}~\bibnamefont
  {Henningson}},\ }\bibfield  {title} {\bibinfo {title} {Direct numerical
  simulation of a separated turbulent boundary layer},\ }\href@noop {}
  {\bibfield  {journal} {\bibinfo  {journal} {J. Fluid Mech.}\ }\textbf
  {\bibinfo {volume} {471}},\ \bibinfo {pages} {107} (\bibinfo {year}
  {2002})}\BibitemShut {NoStop}%
\bibitem [{\citenamefont {Tennekes}\ and\ \citenamefont
  {Lumley}(1972)}]{TennekesLumley1972book}%
  \BibitemOpen
  \bibfield  {author} {\bibinfo {author} {\bibfnamefont {H.}~\bibnamefont
  {Tennekes}}\ and\ \bibinfo {author} {\bibfnamefont {J.~L.}\ \bibnamefont
  {Lumley}},\ }\href@noop {} {\emph {\bibinfo {title} {A First Course In
  Turbulence}}}\ (\bibinfo  {publisher} {The MIT Press},\ \bibinfo {year}
  {1972})\BibitemShut {NoStop}%
\bibitem [{\citenamefont {Sekimoto}\ \emph {et~al.}(2018)\citenamefont
  {Sekimoto}, \citenamefont {Atkinson},\ and\ \citenamefont
  {Soria}}]{SekimotoAtkinsonSoria2018}%
  \BibitemOpen
  \bibfield  {author} {\bibinfo {author} {\bibfnamefont {A.}~\bibnamefont
  {Sekimoto}}, \bibinfo {author} {\bibfnamefont {C.}~\bibnamefont {Atkinson}},\
  and\ \bibinfo {author} {\bibfnamefont {J.}~\bibnamefont {Soria}},\ }\bibfield
   {title} {\bibinfo {title} {Characterisation of minimal-span plane couette
  turbulence with pressure gradients},\ }\href@noop {} {\bibfield  {journal}
  {\bibinfo  {journal} {J. Phys.: Conf. Series}\ }\textbf {\bibinfo {volume}
  {1001}},\ \bibinfo {pages} {012020} (\bibinfo {year} {2018})}\BibitemShut
  {NoStop}%
\bibitem [{\citenamefont {Corrsin}(1958)}]{Corrsin1958}%
  \BibitemOpen
  \bibfield  {author} {\bibinfo {author} {\bibfnamefont {S.}~\bibnamefont
  {Corrsin}},\ }\bibfield  {title} {\bibinfo {title} {Local isotropy in
  turbulent shear flow},\ }\href@noop {} {\bibfield  {journal} {\bibinfo
  {journal} {Res. Memo. NACA.}\ }\textbf {\bibinfo {volume} {58B11}} (\bibinfo
  {year} {1958})}\BibitemShut {NoStop}%
\bibitem [{\citenamefont {Jim\'enez}(2013)}]{Jimenez2013nearwall}%
  \BibitemOpen
  \bibfield  {author} {\bibinfo {author} {\bibfnamefont {J.}~\bibnamefont
  {Jim\'enez}},\ }\bibfield  {title} {\bibinfo {title} {Near-wall
  turbulence.},\ }\href@noop {} {\bibfield  {journal} {\bibinfo  {journal}
  {Phys. Fluids}\ }\textbf {\bibinfo {volume} {25}},\ \bibinfo {pages} {101302}
  (\bibinfo {year} {2013})}\BibitemShut {NoStop}%
\bibitem [{\citenamefont {Sekimoto}\ \emph {et~al.}(2016)\citenamefont
  {Sekimoto}, \citenamefont {Dong},\ and\ \citenamefont
  {Jim\'enez}}]{SekimotoDongJimenez2016}%
  \BibitemOpen
  \bibfield  {author} {\bibinfo {author} {\bibfnamefont {A.}~\bibnamefont
  {Sekimoto}}, \bibinfo {author} {\bibfnamefont {S.}~\bibnamefont {Dong}},\
  and\ \bibinfo {author} {\bibfnamefont {J.}~\bibnamefont {Jim\'enez}},\
  }\bibfield  {title} {\bibinfo {title} {Direct numerical simulation of
  statistically stationary and homogeneous shear turbulence and its relation to
  other shear flows},\ }\href@noop {} {\bibfield  {journal} {\bibinfo
  {journal} {Phys. Fluids}\ }\textbf {\bibinfo {volume} {28}},\ \bibinfo
  {pages} {035101} (\bibinfo {year} {2016})}\BibitemShut {NoStop}%
\bibitem [{\citenamefont {Dong}\ \emph {et~al.}(2017)\citenamefont {Dong},
  \citenamefont {Lozano-Dur\'an}, \citenamefont {Sekimoto},\ and\ \citenamefont
  {Jim\'enez}}]{DongLozanoSekimotoJimenez2017}%
  \BibitemOpen
  \bibfield  {author} {\bibinfo {author} {\bibfnamefont {S.}~\bibnamefont
  {Dong}}, \bibinfo {author} {\bibfnamefont {A.}~\bibnamefont
  {Lozano-Dur\'an}}, \bibinfo {author} {\bibfnamefont {A.}~\bibnamefont
  {Sekimoto}},\ and\ \bibinfo {author} {\bibfnamefont {J.}~\bibnamefont
  {Jim\'enez}},\ }\bibfield  {title} {\bibinfo {title} {Coherent structures in
  statistically stationary homogeneous shear turbulence},\ }\href
  {https://doi.org/10.1017/jfm.2017.78} {\bibfield  {journal} {\bibinfo
  {journal} {Journal of Fluid Mechanics}\ }\textbf {\bibinfo {volume} {816}},\
  \bibinfo {pages} {167–208} (\bibinfo {year} {2017})}\BibitemShut {NoStop}%
\bibitem [{\citenamefont {Borrel}\ and\ \citenamefont
  {Jim\'enez}(2016)}]{BorrelJimenez2016}%
  \BibitemOpen
  \bibfield  {author} {\bibinfo {author} {\bibfnamefont {G.}~\bibnamefont
  {Borrel}}\ and\ \bibinfo {author} {\bibfnamefont {J.}~\bibnamefont
  {Jim\'enez}},\ }\bibfield  {title} {\bibinfo {title} {Properties of the
  turbulent/non-turbulent interface in boundary layers},\ }\href@noop {}
  {\bibfield  {journal} {\bibinfo  {journal} {J. Fluid Mech.}\ }\textbf
  {\bibinfo {volume} {801}},\ \bibinfo {pages} {554} (\bibinfo {year}
  {2016})}\BibitemShut {NoStop}%
\bibitem [{\citenamefont {Sillero}\ \emph {et~al.}(2013)\citenamefont
  {Sillero}, \citenamefont {Jim{\'e}nez},\ and\ \citenamefont
  {Moser}}]{SilleroJimenezMoser2013}%
  \BibitemOpen
  \bibfield  {author} {\bibinfo {author} {\bibfnamefont {J.~A.}\ \bibnamefont
  {Sillero}}, \bibinfo {author} {\bibfnamefont {J.}~\bibnamefont
  {Jim{\'e}nez}},\ and\ \bibinfo {author} {\bibfnamefont {R.~D.}\ \bibnamefont
  {Moser}},\ }\bibfield  {title} {\bibinfo {title} {One-point statistics for
  turbulent wall-bounded flows at {R}eynolds numbers up to {$\delta^+ \approx
  2000$}},\ }\href@noop {} {\bibfield  {journal} {\bibinfo  {journal} {Phys.
  Fluids}\ }\textbf {\bibinfo {volume} {25}},\ \bibinfo {pages} {105102}
  (\bibinfo {year} {2013})}\BibitemShut {NoStop}%
\bibitem [{\citenamefont {Lozano-Dur\'an}\ and\ \citenamefont
  {Jim\'enez}(2014)}]{LozanoJimenez2014pof}%
  \BibitemOpen
  \bibfield  {author} {\bibinfo {author} {\bibfnamefont {A.}~\bibnamefont
  {Lozano-Dur\'an}}\ and\ \bibinfo {author} {\bibfnamefont {J.}~\bibnamefont
  {Jim\'enez}},\ }\bibfield  {title} {\bibinfo {title} {Effect of the
  computational domain on direct simulations of turbulent channels up to
  {$Re_\tau = 4200$}.},\ }\href@noop {} {\bibfield  {journal} {\bibinfo
  {journal} {Phys. Fluids}\ }\textbf {\bibinfo {volume} {26}},\ \bibinfo
  {pages} {011702} (\bibinfo {year} {2014})}\BibitemShut {NoStop}%
\bibitem [{\citenamefont {Simens}\ \emph {et~al.}(2009)\citenamefont {Simens},
  \citenamefont {Jim\'enez}, \citenamefont {Hoyas},\ and\ \citenamefont
  {Mizuno}}]{SimensJimenezHoyasMizuno2009}%
  \BibitemOpen
  \bibfield  {author} {\bibinfo {author} {\bibfnamefont {M.}~\bibnamefont
  {Simens}}, \bibinfo {author} {\bibfnamefont {J.}~\bibnamefont {Jim\'enez}},
  \bibinfo {author} {\bibfnamefont {S.}~\bibnamefont {Hoyas}},\ and\ \bibinfo
  {author} {\bibfnamefont {Y.}~\bibnamefont {Mizuno}},\ }\bibfield  {title}
  {\bibinfo {title} {A high-resolution code for turbulent boundary layers},\
  }\href@noop {} {\bibfield  {journal} {\bibinfo  {journal} {J. Comp. Phys.}\
  }\textbf {\bibinfo {volume} {228}},\ \bibinfo {pages} {4128} (\bibinfo {year}
  {2009})}\BibitemShut {NoStop}%
\bibitem [{\citenamefont {Borrel}\ \emph {et~al.}(2013)\citenamefont {Borrel},
  \citenamefont {Sillero},\ and\ \citenamefont
  {Jim\'enez}}]{BorrelSilleroJimenez2013}%
  \BibitemOpen
  \bibfield  {author} {\bibinfo {author} {\bibfnamefont {G.}~\bibnamefont
  {Borrel}}, \bibinfo {author} {\bibfnamefont {J.~A.}\ \bibnamefont
  {Sillero}},\ and\ \bibinfo {author} {\bibfnamefont {J.}~\bibnamefont
  {Jim\'enez}},\ }\bibfield  {title} {\bibinfo {title} {A code for direct
  numerical simulation of turbulent boundary layers at high {R}eynolds numbers
  in {BG/P} supercomputers},\ }\href@noop {} {\bibfield  {journal} {\bibinfo
  {journal} {Comp. Fluids}\ }\textbf {\bibinfo {volume} {80}},\ \bibinfo
  {pages} {37} (\bibinfo {year} {2013})}\BibitemShut {NoStop}%
\bibitem [{\citenamefont {Clauser}(1954)}]{Clauser1954}%
  \BibitemOpen
  \bibfield  {author} {\bibinfo {author} {\bibfnamefont {F.~H.}\ \bibnamefont
  {Clauser}},\ }\bibfield  {title} {\bibinfo {title} {Turbulent boundary layers
  in adverse pressure gradients},\ }\href@noop {} {\bibfield  {journal}
  {\bibinfo  {journal} {J. Aero. Sci.}\ }\textbf {\bibinfo {volume} {21}},\
  \bibinfo {pages} {91} (\bibinfo {year} {1954})}\BibitemShut {NoStop}%
\bibitem [{\citenamefont {Maciel}\ \emph {et~al.}(2016)\citenamefont {Maciel},
  \citenamefont {Simens},\ and\ \citenamefont
  {Gungor}}]{MacielSimensGungor2016}%
  \BibitemOpen
  \bibfield  {author} {\bibinfo {author} {\bibfnamefont {Y.}~\bibnamefont
  {Maciel}}, \bibinfo {author} {\bibfnamefont {M.~P.}\ \bibnamefont {Simens}},\
  and\ \bibinfo {author} {\bibfnamefont {A.~G.}\ \bibnamefont {Gungor}},\
  }\bibfield  {title} {\bibinfo {title} {Coherent structures in a
  non-equilibrium large-velocity-defect turbulent boundary layer},\ }\href@noop
  {} {\bibfield  {journal} {\bibinfo  {journal} {Flow, Turbulence and
  Combustion}\ ,\ \bibinfo {pages} {1}} (\bibinfo {year} {2016})}\BibitemShut
  {NoStop}%
\bibitem [{\citenamefont {Bradshaw}(1967)}]{Bradshaw1967equilibriumTBL}%
  \BibitemOpen
  \bibfield  {author} {\bibinfo {author} {\bibfnamefont {P.}~\bibnamefont
  {Bradshaw}},\ }\bibfield  {title} {\bibinfo {title} {The turbulence structure
  of equilibrium boundary layers},\ }\href@noop {} {\bibfield  {journal}
  {\bibinfo  {journal} {J. Fluid Mech.}\ }\textbf {\bibinfo {volume} {29}},\
  \bibinfo {pages} {625} (\bibinfo {year} {1967})}\BibitemShut {NoStop}%
\bibitem [{\citenamefont {Wei}\ and\ \citenamefont
  {Maciel}(2018)}]{WeiMaciel2018}%
  \BibitemOpen
  \bibfield  {author} {\bibinfo {author} {\bibfnamefont {T.}~\bibnamefont
  {Wei}}\ and\ \bibinfo {author} {\bibfnamefont {Y.}~\bibnamefont {Maciel}},\
  }\bibfield  {title} {\bibinfo {title} {Derivation of zagarola-smits scaling
  in zero-pressure-gradient turbulent boundary layers},\ }\href@noop {}
  {\bibfield  {journal} {\bibinfo  {journal} {Phys. Rev. Fluids}\ }\textbf
  {\bibinfo {volume} {3}},\ \bibinfo {pages} {012601(R)} (\bibinfo {year}
  {2018})}\BibitemShut {NoStop}%
\bibitem [{\citenamefont {L{\"o}gdberg}\ \emph {et~al.}(2008)\citenamefont
  {L{\"o}gdberg}, \citenamefont {Angele},\ and\ \citenamefont
  {Alfredsson}}]{LogdbergAngeleAlfredsson2008}%
  \BibitemOpen
  \bibfield  {author} {\bibinfo {author} {\bibfnamefont {O.}~\bibnamefont
  {L{\"o}gdberg}}, \bibinfo {author} {\bibfnamefont {K.}~\bibnamefont
  {Angele}},\ and\ \bibinfo {author} {\bibfnamefont {P.~H.}\ \bibnamefont
  {Alfredsson}},\ }\bibfield  {title} {\bibinfo {title} {On the scaling of
  turbulent separating boundary layers},\ }\href@noop {} {\bibfield  {journal}
  {\bibinfo  {journal} {Phys. Fluids}\ }\textbf {\bibinfo {volume} {20}},\
  \bibinfo {pages} {075104} (\bibinfo {year} {2008})}\BibitemShut {NoStop}%
\bibitem [{\citenamefont {Stratford}(1959)}]{Stratford1959}%
  \BibitemOpen
  \bibfield  {author} {\bibinfo {author} {\bibfnamefont {B.}~\bibnamefont
  {Stratford}},\ }\bibfield  {title} {\bibinfo {title} {An experimental flow
  with zero skin friction throughout its region of pressure rise},\ }\href@noop
  {} {\bibfield  {journal} {\bibinfo  {journal} {J. Fluid Mech.}\ }\textbf
  {\bibinfo {volume} {5}},\ \bibinfo {pages} {17} (\bibinfo {year}
  {1959})}\BibitemShut {NoStop}%
\end{thebibliography}%

\end{document}